\documentclass[%
 reprint,
superscriptaddress,
amsmath,amssymb,
aps,
pra,
nolongbibliography
]{revtex4-2}

\usepackage{CJKutf8}
\usepackage{lineno}
\usepackage[colorlinks=true,linkcolor=blue]{hyperref}
\usepackage[squaren]{SIunits}
\usepackage{color}
\usepackage{amsmath}
\usepackage{caption,subcaption}
\usepackage{setspace}
\usepackage{multirow}
\usepackage{threeparttable}
\usepackage{booktabs}
\usepackage{graphicx}
\usepackage{dcolumn}
\usepackage{bm}
\usepackage[T1]{fontenc}

\definecolor{americanrose}{rgb}{1.0, 0.01, 0.24}
\definecolor{coralpink}{rgb}{0.97, 0.51, 0.47}
\definecolor{ao(english)}{rgb}{0.0, 0.5, 0.0}
\definecolor{darkpastelgreen}{rgb}{0.01, 0.75, 0.24}
\definecolor{cyan(process)}{rgb}{0.0, 0.72, 0.92}

\begin{document}

\title{Simulating Vibrationally-Resolved X-ray Photoelectron Spectra of Flexible Molecules: Linear Alkanes C$_{n}$H$_{2n+2}$ ($n$=1-8)}

\author{Xiao Cheng}
 \affiliation{MIIT Key Laboratory of Semiconductor Microstructure and Quantum Sensing, Department of Applied Physics, School of Physics, Nanjing University of Science and Technology, 210094 Nanjing, China}
\affiliation{Hefei National Laboratory for Physical Science at the Microscale, University of Science and Technology of China, 230026 Hefei, China}
\affiliation{Department of Theoretical Chemistry and Biology, School of Engineering Sciences in Chemistry, Biotechnology and Health, Royal Institute of Technology, SE-106 91, Stockholm, Sweden}

\author{Minrui Wei}
\affiliation{MIIT Key Laboratory of Semiconductor Microstructure and Quantum Sensing, Department of Applied Physics, School of Physics, Nanjing University of Science and Technology, 210094 Nanjing, China}

\author{Guangjun Tian}
\affiliation{Key Laboratory for Microstructural Material Physics of Hebei Province, School of Science, Yanshan University, Qinhuangdao 066004, China}


\author{Weijie Hua}
\email{wjhua@njust.edu.cn}
 \affiliation{MIIT Key Laboratory of Semiconductor Microstructure and Quantum Sensing, Department of Applied Physics, School of Physics, Nanjing University of Science and Technology, 210094 Nanjing, China}

\begin{abstract} 
We integrated full core-hole density functional theory with Franck-Condon calculations, considering Duschinsky rotation, to simulate vibrationally-resolved C1s X-ray photoelectron spectra (XPS) of eight linear alkanes, from methane to octane (C$_{n}$H$_{2n+2}$, $n$=1--8). Results align excellently with experimental absolute binding energies and profiles. The spectrum of ethane serves as a ``spectral seed'', with each longer alkane's atom-specific spectrum displaying similar characteristics, albeit with shifts and slight intensity adjustments. Detailed assignments clarify the distinct spectra in short alkanes ($n$=1--4) and their stabilization in long alkanes ($n$=5--8). Carbons are classified as central or distal (C$_1$ and C$_2$), with central carbons contributing nearly identically to the lowest-energy feature A, while distal carbons contribute to the second lowest-energy feature B (both are 0-0 transitions). Increasing molecule size adds more central carbons, enhancing feature A and weakening feature B. Our analysis identifies two dominant Franck-Condon-active vibrations: C$^*$--H stretching ($\sim$3400-3500 cm$^{-1}$) and bending ($\sim$1400-1500 cm$^{-1}$) modes. Structural analysis shows that core ionization minimally affects alkane geometry, less than in ring compounds from previous studies. This work extends our protocol from rigid ring compounds to flexible molecules, contributing to building a high-resolution theoretical XPS library and enhancing the understanding of vibronic coupling.

\end{abstract} 

\maketitle
\section{Introduction}

X-ray photoelectron spectroscopy (XPS) is a powerful analytical technique that provides detailed information about the chemical composition and electronic structure of molecules and materials.\cite{van_der_heide_x-ray_2011} High-resolution XPS spectra allow for investigation of fine structures, which predominantly originate from the vibronic coupling effect. The sub-eV fine structures carry crucial insights into the interplay between nuclear and electronic motions, unveiling the fundamental physics underpinning core hole ionization dynamics and alterations in potential energy surfaces (PESs) between ground and core-ionized states. Historically, vibronic fine structures in XPS spectra were observed as early as 1974 by Siegbahn and coworkers\cite{gelius_vibrational_1974} while studying the spectra of CH$_4$. The continuous advancements in synchrotron radiation technologies have propelled XPS to higher resolutions, unlocking more comprehensive chemical information about the chemical systems.\cite{carravetta_x-ray_2022, hergenhahn_vibrational_2004, svensson_soft_2005, book_ESCA_molecules, ueda_high-resolution_2003, mendolicchio_theory_2019, montorsi_soft_2022, rubensson_vibrationally_2012, quack_high-resolution_2011} 

XPS databases have been helpful in interpreting the structure-spectroscopy relation. However, existing experimental XPS databases\cite{Rumble1992NIST, wagner_1979_handbook, jolly_core-electron_1984, xpsdatabase, sasj, LaSurface} primarily focus on binding energy (BE) values, often omitting intensity data, limiting the depth of analysis. Even for BEs, different experiments sometimes report values that differ over 1 eV. An ultimate goal is to create a comprehensive, high-resolution theoretical library for XPS spectra with both accurate BEs and profiles. Preliminary studies were conducted to investigate a family of molecules (azines\cite{wei_vibronic_2022}, indoles\cite{wei_vibronic_2023}, polycyclic aromatic hydrocarbons (PAHs)\cite{cheng_resolved_2022}, and some other five- and six-membered ring molecules\cite{hua_theoretical_2020, wei_vibronic_2024}, biomolecules\cite{wei_jcp_2024}) with similar structures, by combing density functional theory (DFT) with the full core-hole (FCH) approximation and Franck-Condon (FC) simulations which incorporate the Duschinsky rotation (DR) effect\cite{duschinsky_1937} (denoted as the FCH-DR method for simplicity).\cite{wei_vibronic_2023, wei_jcp_2024, zhanglu_pra_2024} Grounded on the harmonic oscillation (HO) approximation, the calculations yielded good agreement with experiments.\cite{hua_theoretical_2020, wei_vibronic_2022, cheng_resolved_2022, wei_vibronic_2023} The comparative analysis on theoretical and experimental high-resolution spectra offers a more rigorous evaluation, extending beyond solely assessing binding energies, for the accuracy of the chosen theoretical methodology, validating prediction of spectra of related molecules with no experimental data or no high-resolution experimental data. The studies provide  data on equal footing, serving as a reliable stand point for analyzing vibronic properties within a specific family and edge. Extensive investigations were performed including influences on functionals, basis sets, temperature effects, and tautomeric effects, as well as analyses on core-ionization induced structural changes, active vibronal modes, etc. Additionally, the experiment-theory agreement also validated the HO approximation used which is less sensitive for polyatomic molecules than diatomic systems.\cite{zhanglu_pra_2024}

\begin{figure*}
    \includegraphics[width=0.9\textwidth]{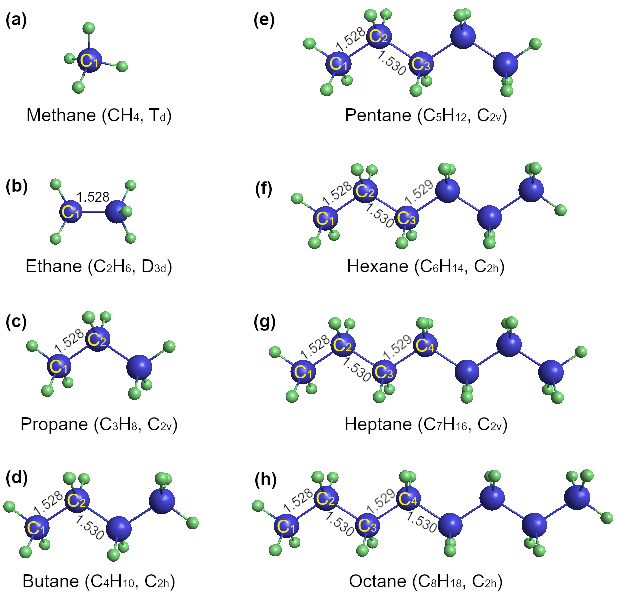}
    \caption{Optimized ground-state structures (i.e., \textbf{min GS}) of the eight linear alkanes under study: (a) methane, (b) ethane, (c) propane, (d) butane, (e) pentane, (f) hextane, (g) heptane, and (h) octane. The chemical formula, symmetry, all non-equivalent C sites, and all C--C bond lengths ({in \AA}) are specified for each molecule.} 
    \label{figure:1:structures}
\end{figure*}

However, all previous calculations on polyatomic systems have predominantly focused on ring compounds which exhibit relatively rigid structures including PAHs. It would be natural to investigate the performance on flexible molecules. In this study, we  extend our research to a series of straight-chain alkanes (C$_{n}$H$_{2n+2}$, $n$=1-8; see Fig. \ref{figure:1:structures}): methane (CH$_4$), ethane (C$_2$H$_6$), propane (C$_3$H$_8$), $n$-butane (C$_4$H$_{10}$), $n$-pentane (C$_5$H$_{12}$), $n$-hexane (C$_6$H$_{14}$), $n$-heptane (C$_7$H$_{16}$), and $n$-octane (C$_8$H$_{18}$).  They represent the simplest organic molecule family, composed solely of \textit{sp}$^3$-hybridized carbon atoms and hydrogen atoms. Gradually increasing the --CH$_2$ groups provide an ideal set for investigating size effects. There are access to high-resolution experimental spectra\cite{karlsen_toward_2002} (except heptane), facilitating the performance evaluation of the theoretical method. Technically, as linear molecules generally possess greater flexibility than their ring counterparts, the geometry optimization of core-ionized states becomes a more intricate task than rigid molecules. 

Alkanes are naturally occurring hydrocarbons found in crude oil, serving as high-value petrochemical feedstock materials for the petroleum industry\cite{stauffer_review_2008}. They also play a prominent role as a major component of non-methane volatile organic compounds in vehicle emissions and urban atmospheric environments\cite{atkinson_atmospheric_2008}. Methane, propane, and butane are clean energy sources for domestic and industrial applications, with methane dominating liquefied natural gas and propane/butane used in liquefied petroleum gas. Ethane, a minor natural gas component, holds significant value as a petrochemical feedstock\cite{xiang_progress_2018, gaffney_ethylene_2017}, primarily for ethylene production. Pentane is used as a polystyrene foam blowing agent and nonpolar solvent, while hexane is a common affordable, safe, and rapidly evaporable non-polar organic solvent. Heptane serves in various industrial applications, including paints, coatings, and pharmaceuticals, and is a significant gasoline component. Octane, another vital gasoline component, is often blended with the highly branched isomer iso-octane to establish a standard reference for measuring the fuel's resistance to engine knock. Methane exhibits T$_\text{d}$ symmetry, while ethane possesses D$_\text{3d}$ symmetry. Longer straight-chain alkanes  have either C$_\text{2v}$ (propane, pentane, and heptane) or C$_\text{2h}$ (butane, hexane, and octane) symmetry, as labeled in Fig. \ref{figure:1:structures} together with all non-equivalent carbon atoms.

It is necessary to acknowledge early theoretical and experimental studies on vibrationally-resolved XPS of alkanes, particularly the work by Svensson and coworkers.\cite{thomas_vibrational_1998, karlsen_toward_2002} They employed equivalent-core hole (ECH) DFT, replacing the core-ionized carbon (C$^*$) with a nitrogen atom,\cite{jolly_thermodynamic_1970, davis_relaxation_1972, plashkevych_validity_2000}  and Franck-Condon simulations to compute spectra for methane, ethane, and other small hydrocarbons, achieving good agreement with experiments.\cite{thomas_vibrational_1998}  Subsequently, they transitioned to another electronic structure method, using a nitrogen basis set for C$^*$ with scaled exponents and an effective core potential (ECP) with adjusted parameters to account for one electron in the C1s shell.\cite{karlsen_toward_2002} This approach was applied to simulate larger linear alkanes ($n=1-6,8$) and extrapolated to alkanes with $n=1000$, providing valuable insights into the vibronic coupling properties of alkanes in XPS.

However, these calculations have several limitations. The ECH-DFT approach\cite{thomas_vibrational_1998} does not yield accurate absolute binding energies. While theoretical spectra can be adjusted to align with experimental values, this method is unsuitable for predictions when no corresponding experiments exist (e.g., heptane). Additionally, our earlier calculations\cite{hua_theoretical_2020} indicated that ECH produces less accurate vibronic profiles compared to FCH, particularly regarding the intensity ratios of the 0-0 and 0-1 peaks, $F_{01}$/$F_{00}$. In the alternative approach\cite{karlsen_toward_2002} using a nitrogen basis set and ECP, scaling issues arise with the basis sets, ECPs, and vibrational frequencies (with different factors for different modes), leading to empirical adjustments that may not be valid for other systems. Furthermore, the linear coupling method (product of Poisson distributions) does not account for mode mixing effects (i.e., Duschinsky rotation). Lastly, while analyses have been conducted on core hole-induced structural changes and potential energy curves, there is a lack of comprehensive examination of active vibrational modes.

The goal of this study is twofold. First, we aim to evaluate the performance of a more accurate FCH-DR approach\cite{hua_theoretical_2020,wei_vibronic_2022,cheng_resolved_2022} for alkane molecules. This computational protocol has been validated\cite{hua_theoretical_2020,wei_vibronic_2022,wei_vibronic_2023, wei_vibronic_2024, cheng_resolved_2022} to accurately predict both binding energies (within 1 eV) and vibronic fine structures for XPS of cyclic compounds, including PAHs. The computational procedure employs only standard basis sets and model core potentials (MCPs) without any empirical parameter scaling. Second, with accurate theoretical spectra at hand, we will investigate general vibronic coupling rules in gaseous linear alkanes. This comprehensive analysis of aliphatic hydrocarbons, alongside our study of PAHs,\cite{cheng_resolved_2022} will enhance our understanding of hydrocarbon systems and contribute to our ongoing goal of constructing a high-resolution XPS library.

\section{Computational methods}

The FCH-DR method\cite{hua_theoretical_2020, cheng_resolved_2022, wei_vibronic_2022} was employed for the XPS computations in this work. Details of the methodology have been outlined previously. Briefly, DFT-based electronic structure calculations were performed using the B3LYP functional\cite{becke_densityfunctional_1993, stephens_ab_1994} and the GAMESS software package\cite{GAMESS}. A double basis set approach\cite{hua_theoretical_2020} was utilized for geometry optimization and frequency calculations, involving distinct basis sets for the ground state ($\chi _{\text{GS}}$) and core-ionized state ($\chi _{\text{FCH}}$). Specifically, the cc-pVTZ basis set was used for $\chi _{\text{GS}}$, while for $\chi _{\text{FCH}}$, the IGLO-III basis set\cite{kutzelnigg1990iglo} was used for core-ionized carbon atoms (C$^*$) and the MCP/NOSeC-V-TZP basis set\cite{sakai_model_1997, noro_contracted_1997, bsjp} was used for non-excited carbon atoms, complemented by the corresponding MCPs. This choice of basis sets has been validated for its accuracy in spectral simulations of various cyclic molecules\cite{hua_theoretical_2020, wei_vibronic_2022}. Fine integration grid and a stringent gradient convergence tolerance of $1.0 \times 10^{-7}$ Hartree/Bohr were used.  The optimized structures and vibrational modes for both the ground and FCH states were used as input to simulate FCFs using our in-house software package, DynaVib\cite{DynaVib},  modified\cite{hua_theoretical_2020} to extend its capabilities from the optical to the X-ray regimes. The DR method\cite{duschinsky_1937} was employed in the computation of FCFs, which characterizes the differences between the normal coordinates of the initial and final states. The $\left\langle 0|n \right\rangle$ amplitudes were computed through an iterative procedure\cite{ruhoff_recursion_1994, ruhoff_algorithms_2000} initiated with $\left\langle 0|0 \right\rangle$. Lorentzian broadening with a half width at half maximum (HWHM) $\gamma$=0.05 eV (i.e., the C1s core hole lifetime\cite{nicolas_lifetime_2012}) was applied to all C1s ionizations, and a small uniform shift $\delta_\text{rel}$=0.2 eV\cite{triguero_separate_1999} was added to account for the scalar relativistic effect. The total XPS profile of each alkane is a weighted combination of the XPS spectrum from its nonequivalent carbon atoms. To better compare with the experimental data,\cite{karlsen_toward_2002} theoretical spectrum of each alkane molecules was shifted by $\delta=$-0.27 eV, except that a slightly different value of $\delta=$-0.33 eV was applied for methane.

\section{Results}

\subsection{Overview and notations for peak labels}

Figure \ref{figure:2:allspec} presents the simulated vibrationally-resolved C1s XPS spectra for all alkanes under study, compared to available experimental data\cite{karlsen_toward_2002} (note that no experimental data is available for heptane). As shown in Figs. \ref{figure:3}--\ref{figure:6}, more detailed analyses--including atom-specific spectra, interpretations of vibronic transitions, and identifications of active modes--are provided for each pair of molecules with the same number of non-equivalent carbons, C$_{n}$H$_{2n+2}$ and C$_{n+1}$H$_{2n+4}$ (where $n=1, 3, 5, 7$).  Figure \ref{figure:7} consolidates the atomic-specific contributions from the edge carbons C$_1$ and C$_2$ to highlight patterns in spectral evolution.

For methane and ethane, each containing only one non-equivalent carbon, we use Roman numerals i--iii and i-v to label their major peaks, respectively. The same numeral indicates the same vibronic assignment and is transferable between the two molecules. For other alkanes with multiple non-equivalent carbons, we use Latin letters A, B, C, etc., to label features in ascending order of energy in the \textit{total} spectrum. These labels are molecule-specific and not necessarily transferable among all molecules, although they are indeed transferable for long alkanes with $n$=5--8.

\subsection{Methane and ethane}

\begin{figure*}
\includegraphics[keepaspectratio]{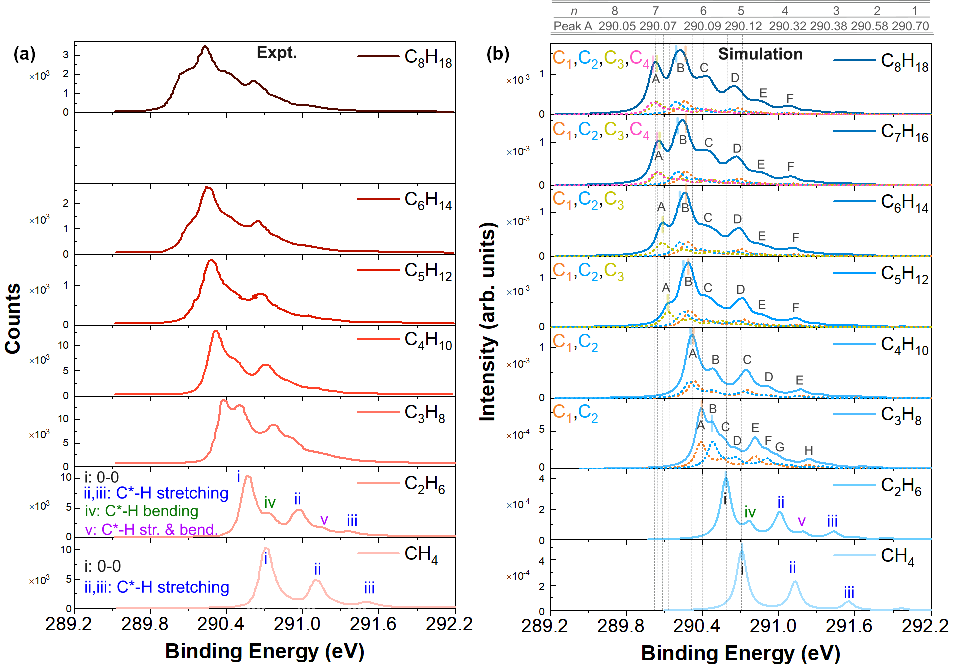}
\caption{Vibrationally-resolved C1s XPS spectra of linear alkane molecules. (a) Experimental spectra recaptured from Karlsen et al.\cite{karlsen_toward_2002} (note that heptane has no available experiment). (b) Our spectral results simulated by using the FCH-DR method.\cite{hua_theoretical_2020,wei_vibronic_2022,cheng_resolved_2022} For methane and ethane with one non-equivalent carbon, major peaks are labeled by Roman numerals and interpreted by active modes (visualized in Fig. \ref{figure:3}). For propane to octane with multiple non-equivalent carbons, atom-specific contributions from carbons C$_n$ (defined in Fig. \ref{figure:1:structures}) are shown with dotted curves, and major peaks in each total spectrum are labeled with Latin letters. Vertical dashed lines (gray) indicate positions of peak A, with values (in eV) given explicitly as a table above panel b. }
\label{figure:2:allspec}
\end{figure*}

\begin{figure*}
\includegraphics[keepaspectratio]{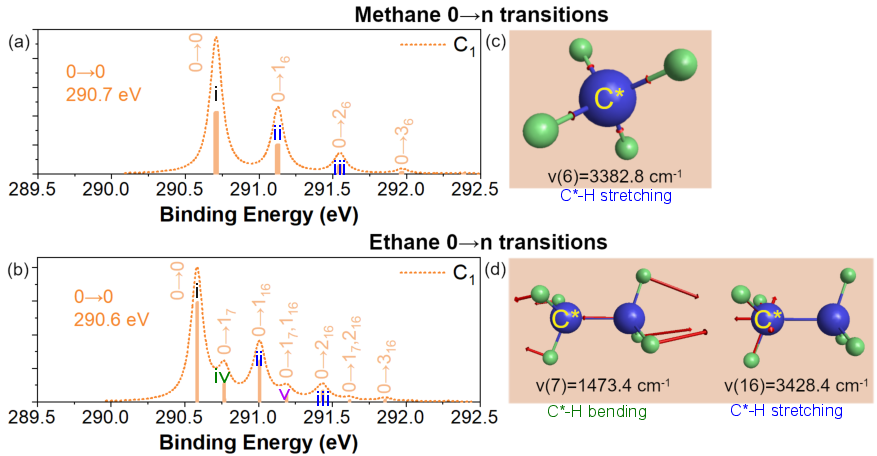}
\caption{Vibrational analysis for methane and ethane. (a,b) Simulated C1s XPS spectra of methane and ethane with vibronic transitions interpreted for major peaks. For instance, in the notation 0$\rightarrow n_{k}$, $n$ stands for the vibrational quantum number and $k$ denotes the mode index. Roman numerals i-v recapture peak labels defined in Fig. \ref{figure:2:allspec}(b). (c,d) Franck-Condon-active vibrational mode(s) interpreted at the optimized fine-state structure (\textbf{min FCH}) of methane and ethane. Main vibrational characteristics near C$^*$ (the core ionization center) is specified under each frequency.} \label{figure:3}
\end{figure*}

The T$_\text{d}$ symmetry molecule methane is the simplest alkane with a single carbon atom. Its experimental spectrum [Fig. \ref{figure:2:allspec}(a)] exhibits three prominent peaks (i-iii) that are well-reproduced by theory [Fig. \ref{figure:2:allspec}(b)]: peaks i (experimental, 290.70 eV; theory, 290.71 eV ), ii (experimental, 291.10 eV; theory, 291.12 eV), and iii (experimental, 291.51 eV; theory, 291.55 eV). One can see that the experiment-theory deviation in peak energies is less than 0.1 eV. These peaks correspond to the 0-0, 0-1, and 0-2 transitions of the totally-symmetric (A$_1$) C$^*$--H stretching mode $\nu(6)$ (3383 cm$^{-1}$) at the final-state geometry \textbf{min FCH}, which is the only Franck-Condon-active mode [Fig. \ref{figure:3}(a,c)].  For clarity, we label peaks ii and iii as 0-1${6}$ and 0-2${6}$, respectively [Fig. \ref{figure:3}(a)], with the mode index indicated as a subscript.\cite{cheng_resolved_2022} Beyond these peaks, no additional features are clearly identifiable in the experimental spectrum [Fig. \ref{figure:2:allspec}(a), bottom], while the 0-3$_{6}$ peak is predicted at approximately 292.00 eV [Fig. \ref{figure:2:allspec}(b), bottom; Fig. \ref{figure:3}(a)].
 
The D$_\text{3d}$ symmetry molecule ethane also contains one single non-equivalent carbon atom. Its theoretical spectrum [Fig. \ref{figure:2:allspec}(b)] closely matches experiment [Fig. \ref{figure:2:allspec}(a)]. The main peak i (0-0 peak) exhibits a red shift of ca. 0.15 eV compared to methane. Peaks ii-iii are again observed, assigned to the C$^*$--H stretching mode $\nu(16)$ (3428 cm$^{-1}$) [Figs. \ref{figure:3}(d), right]. This mode is essentially similar to the only active mode of methane, and the vibrational frequency is slightly (45 cm$^{-1}$) larger than that of methane. Besides, two additional peaks, iv (experimental, 290.73 eV; theory, 290.76 eV) and v (experimental, 291.36 eV; theory, 291.19 eV), are resolved and primarily attributed to the C$^*$--H bending (scissoring) mode $\nu(7)$ (1473 cm$^{-1}$), which also includes contributions from the bending vibrations of the other --CH$_3$ group of the non-excited carbon [Fig. \ref{figure:3}(d), left]. Specifically, peaks iv and v are assigned as the 0$\rightarrow$1$_{7}$ and 0$\rightarrow$1$_{7}$,1$_{16}$ transitions, respectively.

\subsection{Propane and butane}

\begin{figure*}
\includegraphics[width=0.9\textwidth]{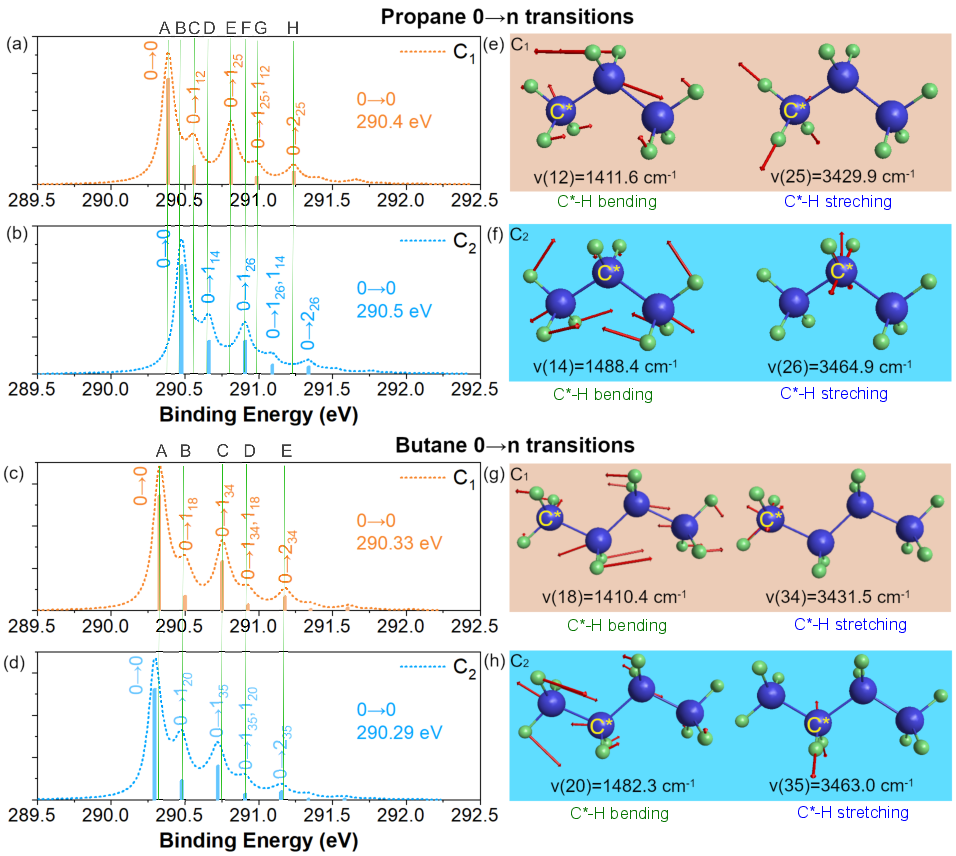}
\caption{
Vibrational analysis for propane and butane. (a--d) Atom-specific contributions of simulated C1s XPS spectra of propane and butane. Underlying vibronic transitions are interpreted for major peaks. Vertical lines (green) marked with  Latin letters A,B, C, $\cdots$ denote  positions of peaks in the total theoretical spectra [Fig. \ref{figure:2:allspec}(b)]. (e--h) Active vibrational modes interpreted at the optimized fine-state structures (\textbf{min FCH}) of propane and butane.  Main characteristics near C$^*$ is specified under each frequency. Major peaks for C$_1$ (labeled as i-v) are interpreted. }
\label{figure:4}
\end{figure*}

Propane with C$_\text{2v}$ symmetry contains two non-equivalent carbons, distal carbon C$_1$ and central carbon C$_2$, in a 2:1 ratio. In the total theoretical spectrum [Fig. \ref{figure:2:allspec}(b)], it now contains rich features (labeled by A-H), since C$_1$ and C$_2$ curves are similar in shape but with different ``phase'' [see also Fig. \ref{figure:4}(a-b)]. Feature A at 290.39 eV and feature B at 290.50 eV are 0-0 transitions by C$_1$  and  C$_2$, respectively. Corresponding experimental values  are 290.40 and 290.50 eV, respectively, almost equivalent to our predictions.  Features C and D come from 0-1$_\text{12}$ and 0-1$_\text{14}$ transitions for C$^*$-H bending modes $\nu(12)$ [1412 cm$^{-1}$, Fig. \ref{figure:4}(e)] for C$_1$ and $\nu(14)$ [1488 cm$^{-1}$, Fig. \ref{figure:4}(f)] for C$_2$, respectively. Meanwhile, peak features E and F come from  0-1$_\text{25}$ and 0-1$_\text{26}$ transitions for C$^*$-H stretching modes $\nu(25)$ [3430 cm$^{-1}$, Fig. \ref{figure:4}(e)] for C$_1$ and $\nu(26)$ [3465 cm$^{-1}$, Fig. \ref{figure:4}(f)] for C$_2$, respectively. Features G and H are both assigned as 0-2 transitions with contributions of both carbons. Feature G is assigned as mixed 0-1 transitions for both the bending and stretching modes; while feature H is assigned as pure 0-2 transitions of the stretching mode.

Similarly, butane with C$_\text{2h}$ symmetry also has two non-equivalent carbons, C$_1$ and C$_2$, in a 2:2 ratio. Compared to propane, the fine structure looks simper because in the C$_1$ and C$_2$ contributions, all peaks almost coincide with each other [see also Fig. \ref{figure:4}(c-d)]. The active mode analyses for butane [Fig. \ref{figure:4}(g-h)] mirror the discussions for propane, identifying a C$^*$-H stretching and a C$^*$-H bending modes for each carbon. Peak A is contributed by  0-0 transitions; peaks B and C are 0-1 transitions for the bending and stretching modes, respectively; peaks D and E are 0-2 transitions, either mixed 0-1 transitions for both the bending and stretching modes (feature D) or pure 0-2 transitions of the stretching mode.

\subsection{Pentane and hextane}

\begin{figure*}
\includegraphics[width=1.0\textwidth]{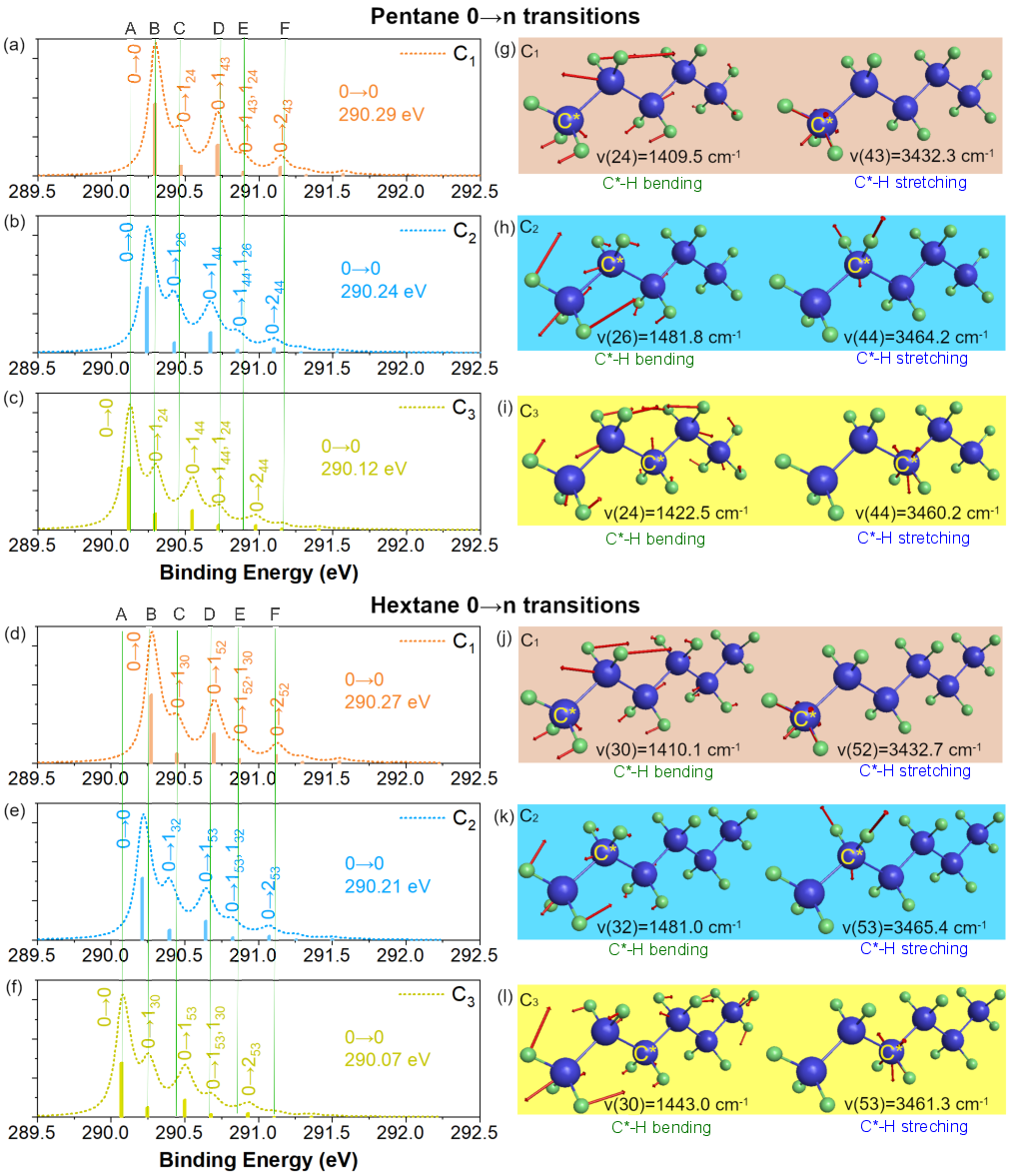}
\caption{Vibrational analysis for pentane and hextane. (a--f) Atom-specific contributions of simulated C1s XPS spectra of pentane and hextane.  Underlying vibronic transitions are interpreted for major peaks. Vertical lines (green) marked with  Latin letters A,B, C, $\cdots$ denote  positions of peaks in the total theoretical spectra [Fig. \ref{figure:2:allspec}(b)]. (g--l) Active vibrational modes interpreted at the optimized fine-state structures  of pentane and hextane.  Main characteristics near C$^*$ is specified under each frequency.}
\label{figure:5}
\end{figure*}

Pentane (C$_\text{2v}$) and hexane (C$_\text{2h}$) both have three non-equivalent carbon atoms, C$_1$--C$_3$, with weight ratios of 2:2:1 and 2:2:2, respectively. Both molecules exhibit similar spectra with six major features A-F. Peak A is assigned as 0-0 transition by the central carbon C$_3$. Peak B is from 0-0 transitions of C$_1$ and C$_2$ as well as 0-1 transition of C$_3$ for a  C$^*$-H bending mode (Fig. \ref{figure:5}). Peaks C and D are assigned as 0-1 transitions, and Peaks E and F as 0-2 transitions. Active modes of the two molecules are analyzed in Fig. \ref{figure:5} (g)-(i) and Fig. \ref{figure:5} (j)-(l). Each atom-specific spectrum is identified with a medium-frequency (1400--1500 cm$^{-1}$) C$^*$--H bending mode and a high-frequency (3400-3500 cm$^{-1}$) C$^*$--H stretching mode.

\subsection{Heptane and octane}

Heptane (C$_\text{2v}$) and octane (C$_\text{2h}$) have four non-equivalent carbon atoms, with weight ratios of 2:2:2:1 and 2:2:2:2, respectively. Although one additional non-equivalent carbon is present as compared to pentane and hextane, the total spectra do not change much, also exhibiting six features A-F. In other words, when $n$ changes from 5 to 8, the vibronic profile almost converges into a consistent pattern. Concerning binding energies, the spectral series exhibit a slight red shift with increasing $n$. Despite the lack of experimental spectrum for heptane, our predicted theoretical spectrum matches the overall evolutionary trend.

\begin{figure*}
\includegraphics[width=1.0\textwidth]{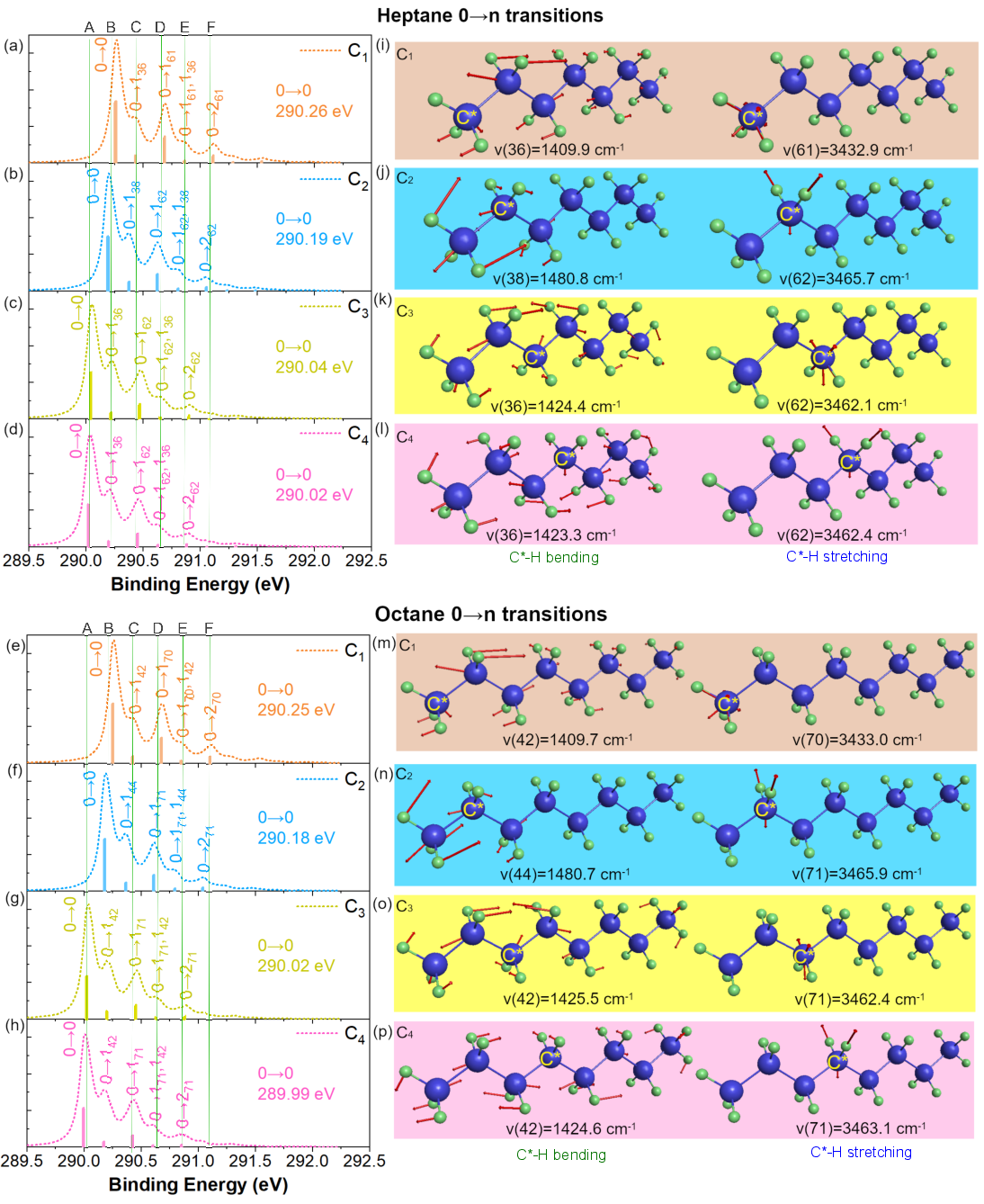}
\caption{Vibrational analysis for heptane and octane. (a--h) Atom-specific contributions of simulated C1s XPS spectra of  heptane and octane.  Underlying vibronic transitions are interpreted for major peaks. Vertical lines (green) marked with  Latin letters A,B, C, $\cdots$ denote  positions of peaks in the total theoretical spectra [Fig. \ref{figure:2:allspec}(b)]. (i--p) Active vibrational modes interpreted at the optimized fine-state structures of heptane and octane. Main vibrational characteristics near C$^*$ is specified.}
\label{figure:6}
\end{figure*}

The underlying reason is that in  heptane and  octane with relatively long length, spectra contributed by the two central carbons C$_3$ and C$_4$ almost coincide with each other, indicating their similar local chemical environment. Their 0-0 transitions both contribute to feature A in the total spectrum. 0-0 transitions of distal carbons C$_1$ and C$_2$ make contributions to feature B some 0.2 eV beyond. The deviation of 0-0 transition energies for C$_1$ and C$_2$ makes peak B broader than peak A. In a even longer linear alkane, it is reasonable to yield that the spectral contributions for two neighboring carbons near the terminal deviates more than two neighboring carbons in the central region.  Based on each atom-specific spectrum  of both molecules [Fig. \ref{figure:6} (a)-(h)], each was identified with two active modes, including one  medium-frequency (1400--1500 cm$^{-1}$) C$^*$--H bending  mode and a high-frequency (3400-3500 cm$^{-1}$) C$^*$--H stretching mode [Fig. \ref{figure:6} (i)-(p)].

\section{Discussion}

\subsection{Red shift of the lowest peak A}

In the total theoretical spectra [Fig. \ref{figure:2:allspec}(b)], the lowest-energy peak, peak A, exhibits a red shift, with energies of 290.70 eV (methane), 290.58 eV (ethane), 290.38 eV (propane), 290.32 eV (butane), 290.12 eV (pentane), 290.09 eV (hexane), 290.07 eV (heptane), and 290.05 eV (octane), as indicated by the numbers at the top of this panel. The magnitude of the red shift is relatively large for alkanes with $n$=1 to 4 but becomes significantly smaller starting from $n$=5. As discussed earlier, for $n$=5 to 8, peak A for each molecule shares the same assignment (from central carbons). The energy differences between neighboring alkanes are only 0.02 to 0.03 eV (approximately 1 milli-Hartree), indicating that the values are approaching convergence.

\subsection{Ratio of distal and central carbons controls intensity ratios of peaks A and B}

Above analyses have shown stable spectral profiles of six features A-F for linear alkanes at $n=5-8$. Peak A and B are 0-0 peaks of central (C$_3$ when $n=5-6$; C$_3$, C$_4$ when $n=7-8$) and distal (C$_1$, C$_2$) carbons, respectively. As $n$ increases, the intensity ratio of peak A to peak B increases due to the increase of more central carbons. This can be explained by the locality of core excitations and the alkane structure (without delocalized $\pi$ structure), which make the atom-specific spectra dominantly determined by the nearest neighboring units. In a long alkane molecule, C$_1$ directly links to a --CH$_2$ group, C$_2$ directly bonds to a -CH$_3$ and --CH$_2$ groups on each side, and all other non-equivalent carbons all bond with two --CH$_2$ groups, which are classified as central carbons. It is worth noting that different central carbons contribute closely similar spectra, as verified by the coincidence of C$_3$ and C$_4$ spectra in heptane and octane. Therefore, in an even longer linear alkane, peak A will be extremely high compared to peak B. The total spectra can be considered as contributed from three types of carbons: C$_1$, C$_2$, and central carbons. This explains the stable features in longer linear alkanes, with their relative peak intensities adjusted by the ratios of different types of carbons and their relative BE differences.

\subsection{Evolution of distal carbons C$_1$ and C$_2$}

\begin{figure*}
\includegraphics[width=0.6\textwidth]{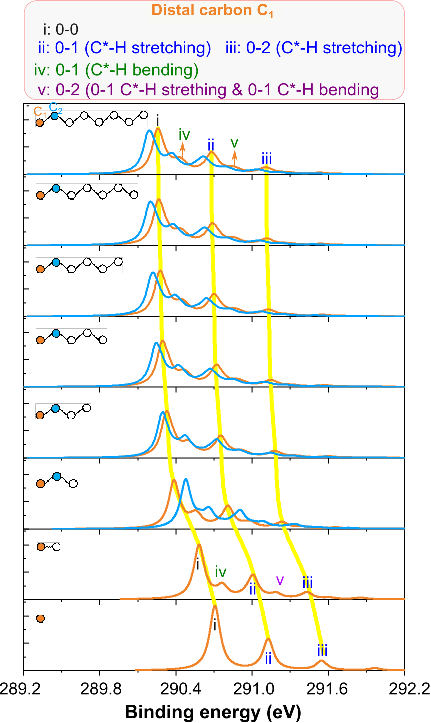}
\caption{
Simulated vibrationally-resolved C1s XPS spectra of distal carbons C$_1$ (orange) and C$_2$ (sky blue) in linear alkanes, showing spectral evolution with molecule length. Peaks i-iii are labeled for methane C$_1$, and peaks i-v for other molecules. Yellow lines highlight selected C$_1$ peak trends. C$_2$ spectra are similar, so no labeling is shown. Molecule structures are schematically depicted in the top left of each panel.}
\label{figure:7}
\end{figure*}

Figure \ref{figure:7} clearly demonstrates the evolution of vibrationally-resolved XPS profiles for the two distal carbon atoms C$_1$ (orange) and C$_2$ (lightblue). The concept of distal carbons is evident for long alkanes with $n$=5--8, while all carbons are considered as distal carbons for short alkanes with $n$=1--4, providing a comprehensive and smooth picture. Methane shows three features i-iii, attributed to the C$^*$--H stretching vibration. Since ethane, an additional C$^*$--H bending vibration is involved, contributing to peaks vi (pure 0-2 bending mode excitation) and v (mixed 0-1 excitations with both modes). All other atom-specific spectral profiles show great similarity to the ethane spectrum, with relative peak intensities of the five peaks preserved almost the same. Figure \ref{figure:7} highlights the evolution of peaks i-iii. These results indicate that the spectrum of ethane can be considered as a ``spectral seed'', as different shifted spectra of ethane approximately generate the total spectra of a longer linear alkane. Additionally, a switch of BE order is found for C$_1$ and C$_2$. Carbon C$_2$ starts since $n$=3, exhibiting a higher binding energy of ca. 0.1 eV than C$_1$. This energy order is quickly reversed since $n$=4, and at $n$=8, the binding energy of C$_2$ is ca. 0.1 eV lower than C$_1$.

\subsection{Structural changes induced by core ionizations}

Structurally, the linear alkanes are generated by gradually adding methylene groups (--CH$_2$--) to methane. Each molecule is constructed by linking C-C and C-H single bonds, and studying the structural change induced by the C1s core hole is crucial. In alkanes, the typical C--C bond length hovers around 1.529 {\AA}, with a tiny difference between distal (1.528 {\AA}) and central carbons (1.529--1.530 {\AA}), as shown in Fig. \ref{figure:1:structures}. The structural changes near the excited carbon are summarized in Table \ref{Table:1}, revealing discernible patterns. Specifically, the bond lengths C$^*$-C$_r$ and C$^*$-C$_l$ in the FCH state exhibit minimal changes (not exceeding 0.02 {\AA}) compared to the ground state structure, indicating marginal influence due to the presence of a C1s core hole. Here, C$_l$ (C$_r$) is the left (right) carbon atom bonded to C$^*$  (see relative atomic positions in Fig. \ref{figure:1:structures}). The angle $\angle$C$_r$-C$^*$-C$_l$, typically close to 114$^\circ$, experiences an increment of ca. 1$^\circ$ from the ground- to the final-state geometry. This angle at C$^*$ experiences slightly larger bending for central carbons compared to distal carbons. All C$^*$--H bonds exhibit virtually identical characteristics in both the ground and the final states, with minor shortening of 0.05--0.07 \AA.

\section{Summary and Conclusions}

To summarize, we simulated high-resolution, vibrationally-resolved C1s X-ray photoelectron spectra (XPS) for a series of linear alkanes, ranging from methane to octane (C$_n$H$_{2n+2}$, with $n$=1--8), by combining full core-hole density functional theory with Franck-Condon calculations that account for Duschinsky rotation. Our theoretical spectra align excellently with all available experimental results, and we predicted the spectra of heptane, showing reasonable spectral evolution in this series. This approach has facilitated reliable peak assignments and enabled an investigation into the spectral evolution with increasing chain length, as well as the underlying vibronic coupling rules governing this molecular family.

The spectrum of ethane serves as a ``spectral seed'', with the vibronic profile of each longer alkane exhibiting similar characteristics, albeit with shifts in peak positions and slight intensity adjustments. The carbon atoms in these longer alkanes are categorized as central or distal (C$_1$ and C$_2$), with central carbons contributing nearly identically to the lowest-energy feature A, while distal carbons contribute to the second-lowest feature B. Detailed assignments elucidate the distinct spectra observed in short alkanes ($n$=1--4) and their stabilization in longer alkanes ($n$=5--8). Through comprehensive assignments, we tracked the evolution of specific spectral features across the molecular series. We investigated core ionization-induced structural changes and identified Franck-Condon-active vibrational modes, highlighting two key active modes present in each atom-specific spectrum: the C$^*$–H symmetric stretching (ca. 3400–3500 cm$^{-1}$) and the C$^*$–H bending (ca. 1400–1500 cm$^{-1}$) modes. 

This work extends our FCH-DR approach to flexible molecular systems, enhancing our understanding of vibronic coupling phenomena. The good performance of our protocol in these flexible systems should benefit from the minimal structural changes induced by core ionization. It also proves that the harmonic oscillator approximation works effectively for alkane systems. The insights gained from this analysis of simple alkane systems will be valuable for studying more complex hydrocarbons and other flexible molecules. Ultimately, this study contributes to our ongoing research goal of constructing a comprehensive, high-resolution theoretical XPS library.

\section*{Acknowledgments}
We thank Prof. Yi Luo for helpful discussions. Financial support from the National Natural Science Foundation of China (Grant No. 12274229)  is greatly acknowledged. M.W. thanks to Fund for Fostering Talented Doctoral Students of Nanjing University of Science and Technology. 

\begin{table*}[]
\begin{threeparttable}
    \centering
    \caption{Local structural changes of each alkane near the core ionization center (C$^*$) at the optimized FCH state structure (\textbf{min FCH}) as compared to its ground state structure (\textbf{min GS}). Bond lengths are in {\AA} and angles in  $^{\circ}$.}
    \renewcommand{\arraystretch}{1.5}
    \setlength{\tabcolsep}{6mm}{
    \begin{tabular}{lccccc}
    \hline
    \hline
    Molecule &  C$^*$ & C$^*$-H & C$^*$-C$_r$\tnote{a} & C$^*$-C$_l$\tnote{a} & $\angle$C$_r$-C$^*$-C$_l$ \\ \hline \hline
    Methane & C$_1$ & 1.037 \hspace{0.2pt} (-0.051)\tnote{b} & ~ & \\
        \hline
    Ethane & ~ & 1.035 \hspace{0.2pt} (-0.056) & 1.526 \hspace{0.2pt} (-0.002) & ~ &   \\ \hline
    \multirow{2}*{Propane} & C$_1$ & 1.034 \hspace{0.2pt} (-0.058) & 1.543 \hspace{0.2pt} (+0.015) & ~ &   \\ 
    ~ & C$_2$ & 1.033 \hspace{0.2pt} (-0.061) & 1.515 \hspace{0.2pt} (-0.013) & \ 1.515 \hspace{0.2pt} (-0.013) & 113.884 \hspace{0.2pt} (+0.878)  \\
        \hline
    \multirow{2}*{Butane} & C$_1$ & 1.034 \hspace{0.2pt} (-0.056) & 1.543 \hspace{0.2pt} (+0.015) & ~ &   \\ 
    ~ & C$_2$ & 1.033 \hspace{0.2pt} (-0.062) & 1.530 \hspace{0.2pt} (+0.001) & 1.512 \hspace{0.2pt} (-0.016) & 114.302 \hspace{0.2pt} (+0.933)  \\ 
        \hline
    \multirow{3}*{Pentane} & C$_1$ & 1.034 \hspace{0.2pt} (-0.058) & 1.544 \hspace{0.2pt} (+0.015) & ~ &   \\
    ~ & C$_2$ & 1.033 \hspace{0.2pt} (-0.062) & 1.530 \hspace{0.2pt} (+0.000) & 1.511 \hspace{0.2pt} (-0.017) & 114.447 \hspace{0.2pt} (+1.109)  \\ 
    ~ & C$_3$ & 1.033 \hspace{0.2pt} (-0.063) & 1.526 \hspace{0.2pt} (-0.003) & 1.526 \hspace{0.2pt} (-0.003) & 114.780 \hspace{0.2pt} (+1.019)  \\ 
        \hline
    \multirow{3}*{Hexane} & C$_1$ & 1.034 \hspace{0.2pt} (-0.058) & 1.544 \hspace{0.2pt} (+0.016) & ~ &   \\ 
    ~ & C$_2$ & 1.033 \hspace{0.2pt} (-0.062) & 1.530 \hspace{0.2pt} (+0.001) & 1.511 \hspace{0.2pt} (-0.017) & 114.483 \hspace{0.2pt} (+1.160)  \\
    ~ & C$_3$ & 1.033 \hspace{0.2pt} (-0.063) & 1.526 \hspace{0.2pt} (-0.004) & 1.526 \hspace{0.2pt} (-0.004) & 114.925 \hspace{0.2pt} (+1.192)  \\
        \hline
    \multirow{4}*{Heptane} & C$_1$ & 1.034 \hspace{0.2pt} (-0.058) & 1.544 \hspace{0.2pt} (+0.016) & ~ &   \\ 
    ~ & C$_2$ & 1.033 \hspace{0.2pt} (-0.062) & 1.531 \hspace{0.2pt} (+0.001) & 1.511 \hspace{0.2pt} (-0.018) & 114.494 \hspace{0.2pt} (+1.168)  \\
    ~ & C$_3$ & 1.033 \hspace{0.2pt} (-0.063) & 1.526 \hspace{0.2pt} (-0.003) & 1.525 \hspace{0.2pt} (-0.005) & 114.957 \hspace{0.2pt} (+1.236)  \\
    ~ & C$_4$ & 1.033 \hspace{0.2pt} (-0.063) & 1.525 \hspace{0.2pt} (-0.004) & 1.525 \hspace{0.2pt} (-0.004) & 115.059 \hspace{0.2pt} (+1.356)  \\
        \hline
    \multirow{4}*{Octane} & C$_1$ & 1.034 \hspace{0.2pt} (-0.058) & 1.545 \hspace{0.2pt} (+0.016) & ~ &   \\ 
    ~ & C$_2$ & 1.033 \hspace{0.2pt} (-0.062) & 1.531 \hspace{0.2pt} (+0.001) & 1.511 \hspace{0.2pt} (-0.018) & 114.511 \hspace{0.2pt} (+1.181)  \\ 
    ~ & C$_3$ & 1.033 \hspace{0.2pt} (-0.063) & 1.527 \hspace{0.2pt} (-0.003) & 1.525 \hspace{0.2pt} (-0.005) & 114.966 \hspace{0.2pt} (+1.241)  \\ 
    ~ & C$_4$ & 1.033 \hspace{0.2pt} (-0.063) & 1.526 \hspace{0.2pt} (-0.004) & 1.524 \hspace{0.2pt} (-0.005) & 115.088 \hspace{0.2pt} (+1.397)  \\ 
        \hline \hline
    \end{tabular}} 
    \label{Table:1}
    \begin{tablenotes}
    \item[a] C$_l$ (C$_r$) is the left (right) carbon atom bonded to C$^*$  (see relative atomic positions in Fig. \ref{figure:1:structures}).
    \item[b] Numbers in parentheses denote the relative values compared with those at the ground state.
    \end{tablenotes}   
\end{threeparttable} 
\end{table*}

\clearpage
\renewcommand{\refname}{References and Notes}

\begin{thebibliography}{46}%
\makeatletter
\providecommand \@ifxundefined [1]{%
 \@ifx{#1\undefined}
}%
\providecommand \@ifnum [1]{%
 \ifnum #1\expandafter \@firstoftwo
 \else \expandafter \@secondoftwo
 \fi
}%
\providecommand \@ifx [1]{%
 \ifx #1\expandafter \@firstoftwo
 \else \expandafter \@secondoftwo
 \fi
}%
\providecommand \natexlab [1]{#1}%
\providecommand \enquote  [1]{``#1''}%
\providecommand \bibnamefont  [1]{#1}%
\providecommand \bibfnamefont [1]{#1}%
\providecommand \citenamefont [1]{#1}%
\providecommand \href@noop [0]{\@secondoftwo}%
\providecommand \href [0]{\begingroup \@sanitize@url \@href}%
\providecommand \@href[1]{\@@startlink{#1}\@@href}%
\providecommand \@@href[1]{\endgroup#1\@@endlink}%
\providecommand \@sanitize@url [0]{\catcode `\\12\catcode `\$12\catcode
  `\&12\catcode `\#12\catcode `\^12\catcode `\_12\catcode `\%12\relax}%
\providecommand \@@startlink[1]{}%
\providecommand \@@endlink[0]{}%
\providecommand \url  [0]{\begingroup\@sanitize@url \@url }%
\providecommand \@url [1]{\endgroup\@href {#1}{\urlprefix }}%
\providecommand \urlprefix  [0]{URL }%
\providecommand \Eprint [0]{\href }%
\providecommand \doibase [0]{https://doi.org/}%
\providecommand \selectlanguage [0]{\@gobble}%
\providecommand \bibinfo  [0]{\@secondoftwo}%
\providecommand \bibfield  [0]{\@secondoftwo}%
\providecommand \translation [1]{[#1]}%
\providecommand \BibitemOpen [0]{}%
\providecommand \bibitemStop [0]{}%
\providecommand \bibitemNoStop [0]{.\EOS\space}%
\providecommand \EOS [0]{\spacefactor3000\relax}%
\providecommand \BibitemShut  [1]{\csname bibitem#1\endcsname}%
\let\auto@bib@innerbib\@empty
\bibitem [{\citenamefont {van~der Heide}(2011)}]{van_der_heide_x-ray_2011}%
  \BibitemOpen
  \bibfield  {author} {\bibinfo {author} {\bibfnamefont {P.}~\bibnamefont
  {van~der Heide}},\ }\href {https://doi.org/10.1002/9781118162897} {\emph
  {\bibinfo {title} {X-{Ray} {Photoelectron} {Spectroscopy}: {An}
  {Introduction} to {Principles} and {Practices}}}}\ (\bibinfo  {publisher}
  {John Wiley \& Sons, Inc.},\ \bibinfo {address} {Hoboken, NJ, USA},\ \bibinfo
  {year} {2011})\BibitemShut {NoStop}%
\bibitem [{\citenamefont {Gelius}\ \emph {et~al.}(1974)\citenamefont {Gelius},
  \citenamefont {Svensson}, \citenamefont {Siegbahn}, \citenamefont {Basilier},
  \citenamefont {Fax{\"a}lv},\ and\ \citenamefont
  {Siegbahn}}]{gelius_vibrational_1974}%
  \BibitemOpen
  \bibfield  {author} {\bibinfo {author} {\bibfnamefont {U.}~\bibnamefont
  {Gelius}}, \bibinfo {author} {\bibfnamefont {S.}~\bibnamefont {Svensson}},
  \bibinfo {author} {\bibfnamefont {H.}~\bibnamefont {Siegbahn}}, \bibinfo
  {author} {\bibfnamefont {E.}~\bibnamefont {Basilier}}, \bibinfo {author}
  {\bibfnamefont {{\r A}.}~\bibnamefont {Fax{\"a}lv}},\ and\ \bibinfo {author}
  {\bibfnamefont {K.}~\bibnamefont {Siegbahn}},\ }\href@noop {} {\bibfield
  {journal} {\bibinfo  {journal} {Chem. Phys. Lett.}\ }\textbf {\bibinfo
  {volume} {28}},\ \bibinfo {pages} {1} (\bibinfo {year} {1974})}\BibitemShut
  {NoStop}%
\bibitem [{\citenamefont {Carravetta}\ \emph {et~al.}(2022)\citenamefont
  {Carravetta}, \citenamefont {Couto},\ and\ \citenamefont
  {{\AA}gren}}]{carravetta_x-ray_2022}%
  \BibitemOpen
  \bibfield  {author} {\bibinfo {author} {\bibfnamefont {V.}~\bibnamefont
  {Carravetta}}, \bibinfo {author} {\bibfnamefont {R.~C.}\ \bibnamefont
  {Couto}},\ and\ \bibinfo {author} {\bibfnamefont {H.}~\bibnamefont
  {{\AA}gren}},\ }\href {https://doi.org/10.1088/1361-648X/ac7d2a} {\bibfield
  {journal} {\bibinfo  {journal} {J. Phys.: Condens. Matter}\ }\textbf
  {\bibinfo {volume} {34}},\ \bibinfo {pages} {363002} (\bibinfo {year}
  {2022})}\BibitemShut {NoStop}%
\bibitem [{\citenamefont {Hergenhahn}(2004)}]{hergenhahn_vibrational_2004}%
  \BibitemOpen
  \bibfield  {author} {\bibinfo {author} {\bibfnamefont {U.}~\bibnamefont
  {Hergenhahn}},\ }\href {https://doi.org/10.1088/0953-4075/37/12/R01}
  {\bibfield  {journal} {\bibinfo  {journal} {J. Phys. B: At. Mol. Opt. Phys.}\
  }\textbf {\bibinfo {volume} {37}},\ \bibinfo {pages} {R89} (\bibinfo {year}
  {2004})}\BibitemShut {NoStop}%
\bibitem [{\citenamefont {Svensson}(2005)}]{svensson_soft_2005}%
  \BibitemOpen
  \bibfield  {author} {\bibinfo {author} {\bibfnamefont {S.}~\bibnamefont
  {Svensson}},\ }\href {https://doi.org/10.1088/0953-4075/38/9/024} {\bibfield
  {journal} {\bibinfo  {journal} {J. Phys. B: At. Mol. Opt. Phys.}\ }\textbf
  {\bibinfo {volume} {38}},\ \bibinfo {pages} {S821} (\bibinfo {year}
  {2005})}\BibitemShut {NoStop}%
\bibitem [{\citenamefont {Siegbahn}(1969)}]{book_ESCA_molecules}%
  \BibitemOpen
  \bibfield  {author} {\bibinfo {author} {\bibfnamefont {K.}~\bibnamefont
  {Siegbahn}},\ }\href@noop {} {\emph {\bibinfo {title} {ESCA applied to free
  molecules}}}\ (\bibinfo  {publisher} {North-Holland Publishing},\ \bibinfo
  {year} {1969})\BibitemShut {NoStop}%
\bibitem [{\citenamefont {Ueda}(2003)}]{ueda_high-resolution_2003}%
  \BibitemOpen
  \bibfield  {author} {\bibinfo {author} {\bibfnamefont {K.}~\bibnamefont
  {Ueda}},\ }\href {https://doi.org/10.1088/0953-4075/36/4/201} {\bibfield
  {journal} {\bibinfo  {journal} {J. Phys. B: At. Mol. Opt. Phys.}\ }\textbf
  {\bibinfo {volume} {36}},\ \bibinfo {pages} {R1} (\bibinfo {year}
  {2003})}\BibitemShut {NoStop}%
\bibitem [{\citenamefont {Mendolicchio}\ \emph {et~al.}(2019)\citenamefont
  {Mendolicchio}, \citenamefont {Baiardi}, \citenamefont {Fronzoni},
  \citenamefont {Stener}, \citenamefont {Grazioli}, \citenamefont {de~Simone},\
  and\ \citenamefont {Barone}}]{mendolicchio_theory_2019}%
  \BibitemOpen
  \bibfield  {author} {\bibinfo {author} {\bibfnamefont {M.}~\bibnamefont
  {Mendolicchio}}, \bibinfo {author} {\bibfnamefont {A.}~\bibnamefont
  {Baiardi}}, \bibinfo {author} {\bibfnamefont {G.}~\bibnamefont {Fronzoni}},
  \bibinfo {author} {\bibfnamefont {M.}~\bibnamefont {Stener}}, \bibinfo
  {author} {\bibfnamefont {C.}~\bibnamefont {Grazioli}}, \bibinfo {author}
  {\bibfnamefont {M.}~\bibnamefont {de~Simone}},\ and\ \bibinfo {author}
  {\bibfnamefont {V.}~\bibnamefont {Barone}},\ }\href
  {https://doi.org/10.1063/1.5122310} {\bibfield  {journal} {\bibinfo
  {journal} {J. Chem. Phys.}\ }\textbf {\bibinfo {volume} {151}},\ \bibinfo
  {pages} {124105} (\bibinfo {year} {2019})}\BibitemShut {NoStop}%
\bibitem [{\citenamefont {Montorsi}\ \emph {et~al.}(2022)\citenamefont
  {Montorsi}, \citenamefont {Segatta}, \citenamefont {Nenov}, \citenamefont
  {Mukamel},\ and\ \citenamefont {Garavelli}}]{montorsi_soft_2022}%
  \BibitemOpen
  \bibfield  {author} {\bibinfo {author} {\bibfnamefont {F.}~\bibnamefont
  {Montorsi}}, \bibinfo {author} {\bibfnamefont {F.}~\bibnamefont {Segatta}},
  \bibinfo {author} {\bibfnamefont {A.}~\bibnamefont {Nenov}}, \bibinfo
  {author} {\bibfnamefont {S.}~\bibnamefont {Mukamel}},\ and\ \bibinfo {author}
  {\bibfnamefont {M.}~\bibnamefont {Garavelli}},\ }\href
  {https://doi.org/10.1021/acs.jctc.1c00566} {\bibfield  {journal} {\bibinfo
  {journal} {J. Chem. Theory Comput.}\ }\textbf {\bibinfo {volume} {18}},\
  \bibinfo {pages} {1003} (\bibinfo {year} {2022})}\BibitemShut {NoStop}%
\bibitem [{\citenamefont {Rubensson}\ \emph {et~al.}(2012)\citenamefont
  {Rubensson}, \citenamefont {Pietzsch},\ and\ \citenamefont
  {Hennies}}]{rubensson_vibrationally_2012}%
  \BibitemOpen
  \bibfield  {author} {\bibinfo {author} {\bibfnamefont {J.-E.}\ \bibnamefont
  {Rubensson}}, \bibinfo {author} {\bibfnamefont {A.}~\bibnamefont
  {Pietzsch}},\ and\ \bibinfo {author} {\bibfnamefont {F.}~\bibnamefont
  {Hennies}},\ }\href {https://doi.org/10.1016/j.elspec.2012.05.003} {\bibfield
   {journal} {\bibinfo  {journal} {J. Electron Spectros. Relat. Phenomena}\
  }\textbf {\bibinfo {volume} {185}},\ \bibinfo {pages} {294} (\bibinfo {year}
  {2012})}\BibitemShut {NoStop}%
\bibitem [{\citenamefont {Miron}\ and\ \citenamefont
  {Morin}(2011)}]{quack_high-resolution_2011}%
  \BibitemOpen
  \bibfield  {author} {\bibinfo {author} {\bibfnamefont {C.}~\bibnamefont
  {Miron}}\ and\ \bibinfo {author} {\bibfnamefont {P.}~\bibnamefont {Morin}},\
  }in\ \href {https://doi.org/10.1002/9780470749593.hrs066} {\emph {\bibinfo
  {booktitle} {Handbook of {High}-resolution {Spectroscopy}}}},\ \bibinfo
  {editor} {edited by\ \bibinfo {editor} {\bibfnamefont {M.}~\bibnamefont
  {Quack}}\ and\ \bibinfo {editor} {\bibfnamefont {F.}~\bibnamefont {Merkt}}}\
  (\bibinfo  {publisher} {John Wiley \& Sons, Ltd},\ \bibinfo {address}
  {Chichester, UK},\ \bibinfo {year} {2011})\ p.\ \bibinfo {pages}
  {hrs066}\BibitemShut {NoStop}%
\bibitem [{\citenamefont {Rumble~Jr.}\ \emph {et~al.}(1992)\citenamefont
  {Rumble~Jr.}, \citenamefont {Bickham},\ and\ \citenamefont
  {Powell}}]{Rumble1992NIST}%
  \BibitemOpen
  \bibfield  {author} {\bibinfo {author} {\bibfnamefont {J.~R.}\ \bibnamefont
  {Rumble~Jr.}}, \bibinfo {author} {\bibfnamefont {D.~M.}\ \bibnamefont
  {Bickham}},\ and\ \bibinfo {author} {\bibfnamefont {C.~J.}\ \bibnamefont
  {Powell}},\ }\href@noop {} {\bibfield  {journal} {\bibinfo  {journal} {Surf.
  Interface Anal.}\ }\textbf {\bibinfo {volume} {19}},\ \bibinfo {pages} {241}
  (\bibinfo {year} {1992})}\BibitemShut {NoStop}%
\bibitem [{\citenamefont {Wagner}(1979)}]{wagner_1979_handbook}%
  \BibitemOpen
  \bibfield  {author} {\bibinfo {author} {\bibfnamefont {C.~D.}\ \bibnamefont
  {Wagner}},\ }\href@noop {} {\emph {\bibinfo {title} {Handbook of x-ray
  photoelectron spectroscopy: a reference book of standard data for use in
  x-ray photoelectron spectroscopy}}}\ (\bibinfo  {publisher} {Perkin-Elmer},\
  \bibinfo {address} {Shelton, CT},\ \bibinfo {year} {1979})\BibitemShut
  {NoStop}%
\bibitem [{\citenamefont {Jolly}\ \emph {et~al.}(1984)\citenamefont {Jolly},
  \citenamefont {Bomben},\ and\ \citenamefont
  {Eyermann}}]{jolly_core-electron_1984}%
  \BibitemOpen
  \bibfield  {author} {\bibinfo {author} {\bibfnamefont {W.}~\bibnamefont
  {Jolly}}, \bibinfo {author} {\bibfnamefont {K.}~\bibnamefont {Bomben}},\ and\
  \bibinfo {author} {\bibfnamefont {C.}~\bibnamefont {Eyermann}},\ }\href@noop
  {} {\bibfield  {journal} {\bibinfo  {journal} {Atom. Data Nucl. Data}\
  }\textbf {\bibinfo {volume} {31}},\ \bibinfo {pages} {433} (\bibinfo {year}
  {1984})}\BibitemShut {NoStop}%
\bibitem [{xps()}]{xpsdatabase}%
  \BibitemOpen
  \href@noop {} {}\bibinfo {note} {Https://xpsdatabase.com/, Accessed on
  2024-7-20.}\BibitemShut {Stop}%
\bibitem [{sas()}]{sasj}%
  \BibitemOpen
  \href@noop {} {}\bibinfo {note} {Http://www.sasj.jp/, Accessed on
  2024-7-20.}\BibitemShut {Stop}%
\bibitem [{LaS()}]{LaSurface}%
  \BibitemOpen
  \href@noop {} {}\bibinfo {note} {Http://www.lasurface.com/xps/index.php/,
  Accessed on 2024-7-20.}\BibitemShut {Stop}%
\bibitem [{\citenamefont {Wei}\ \emph {et~al.}(2022)\citenamefont {Wei},
  \citenamefont {Cheng}, \citenamefont {Zhang}, \citenamefont {Zhang},
  \citenamefont {Wang}, \citenamefont {Ge}, \citenamefont {Tian},\ and\
  \citenamefont {Hua}}]{wei_vibronic_2022}%
  \BibitemOpen
  \bibfield  {author} {\bibinfo {author} {\bibfnamefont {M.}~\bibnamefont
  {Wei}}, \bibinfo {author} {\bibfnamefont {X.}~\bibnamefont {Cheng}}, \bibinfo
  {author} {\bibfnamefont {L.}~\bibnamefont {Zhang}}, \bibinfo {author}
  {\bibfnamefont {J.-R.}\ \bibnamefont {Zhang}}, \bibinfo {author}
  {\bibfnamefont {S.-Y.}\ \bibnamefont {Wang}}, \bibinfo {author}
  {\bibfnamefont {G.}~\bibnamefont {Ge}}, \bibinfo {author} {\bibfnamefont
  {G.}~\bibnamefont {Tian}},\ and\ \bibinfo {author} {\bibfnamefont
  {W.}~\bibnamefont {Hua}},\ }\href
  {https://doi.org/10.1103/PhysRevA.106.022811} {\bibfield  {journal} {\bibinfo
   {journal} {Phys. Rev. A}\ }\textbf {\bibinfo {volume} {106}},\ \bibinfo
  {pages} {022811} (\bibinfo {year} {2022})}\BibitemShut {NoStop}%
\bibitem [{\citenamefont {Wei}\ \emph {et~al.}(2023)\citenamefont {Wei},
  \citenamefont {Zhang}, \citenamefont {Tian},\ and\ \citenamefont
  {Hua}}]{wei_vibronic_2023}%
  \BibitemOpen
  \bibfield  {author} {\bibinfo {author} {\bibfnamefont {M.}~\bibnamefont
  {Wei}}, \bibinfo {author} {\bibfnamefont {L.}~\bibnamefont {Zhang}}, \bibinfo
  {author} {\bibfnamefont {G.}~\bibnamefont {Tian}},\ and\ \bibinfo {author}
  {\bibfnamefont {W.}~\bibnamefont {Hua}},\ }\href@noop {} {\bibfield
  {journal} {\bibinfo  {journal} {Phys. Rev. A}\ }\textbf {\bibinfo {volume}
  {108}},\ \bibinfo {pages} {022816} (\bibinfo {year} {2023})}\BibitemShut
  {NoStop}%
\bibitem [{\citenamefont {Cheng}\ \emph {et~al.}(2022)\citenamefont {Cheng},
  \citenamefont {Wei}, \citenamefont {Tian}, \citenamefont {Luo},\ and\
  \citenamefont {Hua}}]{cheng_resolved_2022}%
  \BibitemOpen
  \bibfield  {author} {\bibinfo {author} {\bibfnamefont {X.}~\bibnamefont
  {Cheng}}, \bibinfo {author} {\bibfnamefont {M.}~\bibnamefont {Wei}}, \bibinfo
  {author} {\bibfnamefont {G.}~\bibnamefont {Tian}}, \bibinfo {author}
  {\bibfnamefont {Y.}~\bibnamefont {Luo}},\ and\ \bibinfo {author}
  {\bibfnamefont {W.}~\bibnamefont {Hua}},\ }\href
  {https://doi.org/10.1021/acs.jpca.2c04426} {\bibfield  {journal} {\bibinfo
  {journal} {J. Phys. Chem. A}\ }\textbf {\bibinfo {volume} {126}},\ \bibinfo
  {pages} {5582} (\bibinfo {year} {2022})}\BibitemShut {NoStop}%
\bibitem [{\citenamefont {Hua}\ \emph {et~al.}(2020)\citenamefont {Hua},
  \citenamefont {Tian},\ and\ \citenamefont {Luo}}]{hua_theoretical_2020}%
  \BibitemOpen
  \bibfield  {author} {\bibinfo {author} {\bibfnamefont {W.}~\bibnamefont
  {Hua}}, \bibinfo {author} {\bibfnamefont {G.}~\bibnamefont {Tian}},\ and\
  \bibinfo {author} {\bibfnamefont {Y.}~\bibnamefont {Luo}},\ }\href
  {https://doi.org/10.1039/D0CP02970J} {\bibfield  {journal} {\bibinfo
  {journal} {Phys. Chem. Chem. Phys.}\ }\textbf {\bibinfo {volume} {22}},\
  \bibinfo {pages} {20014} (\bibinfo {year} {2020})}\BibitemShut {NoStop}%
\bibitem [{\citenamefont {Wei}\ \emph {et~al.}(2024{\natexlab{a}})\citenamefont
  {Wei}, \citenamefont {Zuo}, \citenamefont {Tian},\ and\ \citenamefont
  {Hua}}]{wei_vibronic_2024}%
  \BibitemOpen
  \bibfield  {author} {\bibinfo {author} {\bibfnamefont {M.}~\bibnamefont
  {Wei}}, \bibinfo {author} {\bibfnamefont {J.}~\bibnamefont {Zuo}}, \bibinfo
  {author} {\bibfnamefont {G.}~\bibnamefont {Tian}},\ and\ \bibinfo {author}
  {\bibfnamefont {W.}~\bibnamefont {Hua}},\ }\href
  {https://doi.org/10.1103/PhysRevA.109.022820} {\bibfield  {journal} {\bibinfo
   {journal} {Phys. Rev. A}\ }\textbf {\bibinfo {volume} {109}},\ \bibinfo
  {pages} {022820} (\bibinfo {year} {2024}{\natexlab{a}})}\BibitemShut
  {NoStop}%
\bibitem [{\citenamefont {Wei}\ \emph {et~al.}(2024{\natexlab{b}})\citenamefont
  {Wei}, \citenamefont {Zuo}, \citenamefont {Tian},\ and\ \citenamefont
  {Hua}}]{wei_jcp_2024}%
  \BibitemOpen
  \bibfield  {author} {\bibinfo {author} {\bibfnamefont {M.}~\bibnamefont
  {Wei}}, \bibinfo {author} {\bibfnamefont {J.}~\bibnamefont {Zuo}}, \bibinfo
  {author} {\bibfnamefont {G.}~\bibnamefont {Tian}},\ and\ \bibinfo {author}
  {\bibfnamefont {W.}~\bibnamefont {Hua}},\ }\href@noop {} {\bibfield
  {journal} {\bibinfo  {journal} {arXiv preprint arXiv:2403.09109}\ } (\bibinfo
  {year} {2024}{\natexlab{b}})}\BibitemShut {NoStop}%
\bibitem [{\citenamefont {Duschinsky}(1937)}]{duschinsky_1937}%
  \BibitemOpen
  \bibfield  {author} {\bibinfo {author} {\bibfnamefont {F.}~\bibnamefont
  {Duschinsky}},\ }\href@noop {} {\bibfield  {journal} {\bibinfo  {journal}
  {Acta Physicochim. URSS}\ }\textbf {\bibinfo {volume} {7}},\ \bibinfo {pages}
  {551} (\bibinfo {year} {1937})},\ \bibinfo {note} {and its English
  translation by C. W. M\"{u}ler available from
  http://www.chem.purdue.edu/zwier/pubs/Duschinsky.pdf}\BibitemShut {NoStop}%
\bibitem [{\citenamefont {Zhang}\ \emph {et~al.}(2024)\citenamefont {Zhang},
  \citenamefont {Wei}, \citenamefont {Ge},\ and\ \citenamefont
  {Hua}}]{zhanglu_pra_2024}%
  \BibitemOpen
  \bibfield  {author} {\bibinfo {author} {\bibfnamefont {L.}~\bibnamefont
  {Zhang}}, \bibinfo {author} {\bibfnamefont {M.}~\bibnamefont {Wei}}, \bibinfo
  {author} {\bibfnamefont {G.}~\bibnamefont {Ge}},\ and\ \bibinfo {author}
  {\bibfnamefont {W.}~\bibnamefont {Hua}},\ }\href
  {https://doi.org/10.1103/PhysRevA.109.032815} {\bibfield  {journal} {\bibinfo
   {journal} {Phys. Rev. A}\ }\textbf {\bibinfo {volume} {109}},\ \bibinfo
  {pages} {032815} (\bibinfo {year} {2024})}\BibitemShut {NoStop}%
\bibitem [{\citenamefont {Karlsen}\ \emph {et~al.}(2002)\citenamefont
  {Karlsen}, \citenamefont {Børve}, \citenamefont {Sæthre}, \citenamefont
  {Wiesner}, \citenamefont {Bässler},\ and\ \citenamefont
  {Svensson}}]{karlsen_toward_2002}%
  \BibitemOpen
  \bibfield  {author} {\bibinfo {author} {\bibfnamefont {T.}~\bibnamefont
  {Karlsen}}, \bibinfo {author} {\bibfnamefont {K.~J.}\ \bibnamefont {Børve}},
  \bibinfo {author} {\bibfnamefont {L.~J.}\ \bibnamefont {Sæthre}}, \bibinfo
  {author} {\bibfnamefont {K.}~\bibnamefont {Wiesner}}, \bibinfo {author}
  {\bibfnamefont {M.}~\bibnamefont {Bässler}},\ and\ \bibinfo {author}
  {\bibfnamefont {S.}~\bibnamefont {Svensson}},\ }\href
  {https://doi.org/10.1021/ja010649j} {\bibfield  {journal} {\bibinfo
  {journal} {J. Am. Chem. Soc.}\ }\textbf {\bibinfo {volume} {124}},\ \bibinfo
  {pages} {7866} (\bibinfo {year} {2002})}\BibitemShut {NoStop}%
\bibitem [{\citenamefont {Stauffer}\ \emph {et~al.}(2008)\citenamefont
  {Stauffer}, \citenamefont {Dolan},\ and\ \citenamefont
  {Newman}}]{stauffer_review_2008}%
  \BibitemOpen
  \bibfield  {author} {\bibinfo {author} {\bibfnamefont {E.}~\bibnamefont
  {Stauffer}}, \bibinfo {author} {\bibfnamefont {J.~A.}\ \bibnamefont
  {Dolan}},\ and\ \bibinfo {author} {\bibfnamefont {R.}~\bibnamefont
  {Newman}},\ }in\ \href {https://doi.org/10.1016/B978-012663971-1.50007-5}
  {\emph {\bibinfo {booktitle} {Fire {Debris} {Analysis}}}}\ (\bibinfo
  {publisher} {Elsevier},\ \bibinfo {year} {2008})\ pp.\ \bibinfo {pages}
  {49--83}\BibitemShut {NoStop}%
\bibitem [{\citenamefont {Atkinson}\ \emph {et~al.}(2008)\citenamefont
  {Atkinson}, \citenamefont {Arey},\ and\ \citenamefont
  {Aschmann}}]{atkinson_atmospheric_2008}%
  \BibitemOpen
  \bibfield  {author} {\bibinfo {author} {\bibfnamefont {R.}~\bibnamefont
  {Atkinson}}, \bibinfo {author} {\bibfnamefont {J.}~\bibnamefont {Arey}},\
  and\ \bibinfo {author} {\bibfnamefont {S.~M.}\ \bibnamefont {Aschmann}},\
  }\href {https://doi.org/https://doi.org/10.1016/j.atmosenv.2007.08.040}
  {\bibfield  {journal} {\bibinfo  {journal} {Atmos. Environ.}\ }\textbf
  {\bibinfo {volume} {42}},\ \bibinfo {pages} {5859} (\bibinfo {year}
  {2008})}\BibitemShut {NoStop}%
\bibitem [{\citenamefont {Xiang}\ \emph {et~al.}(2018)\citenamefont {Xiang},
  \citenamefont {Wang}, \citenamefont {Cheng},\ and\ \citenamefont
  {Matsubu}}]{xiang_progress_2018}%
  \BibitemOpen
  \bibfield  {author} {\bibinfo {author} {\bibfnamefont {Y.}~\bibnamefont
  {Xiang}}, \bibinfo {author} {\bibfnamefont {H.}~\bibnamefont {Wang}},
  \bibinfo {author} {\bibfnamefont {J.}~\bibnamefont {Cheng}},\ and\ \bibinfo
  {author} {\bibfnamefont {J.}~\bibnamefont {Matsubu}},\ }\href
  {https://doi.org/10.1039/C7CY01878A} {\bibfield  {journal} {\bibinfo
  {journal} {Catal. Sci. Technol.}\ }\textbf {\bibinfo {volume} {8}},\ \bibinfo
  {pages} {1500} (\bibinfo {year} {2018})}\BibitemShut {NoStop}%
\bibitem [{\citenamefont {Gaffney}\ and\ \citenamefont
  {Mason}(2017)}]{gaffney_ethylene_2017}%
  \BibitemOpen
  \bibfield  {author} {\bibinfo {author} {\bibfnamefont {A.~M.}\ \bibnamefont
  {Gaffney}}\ and\ \bibinfo {author} {\bibfnamefont {O.~M.}\ \bibnamefont
  {Mason}},\ }\href {https://doi.org/10.1016/j.cattod.2017.01.020} {\bibfield
  {journal} {\bibinfo  {journal} {Catal. Today}\ }\textbf {\bibinfo {volume}
  {285}},\ \bibinfo {pages} {159} (\bibinfo {year} {2017})}\BibitemShut
  {NoStop}%
\bibitem [{\citenamefont {Thomas}\ \emph {et~al.}(1998)\citenamefont {Thomas},
  \citenamefont {Saethre}, \citenamefont {Sorensen},\ and\ \citenamefont
  {Svensson}}]{thomas_vibrational_1998}%
  \BibitemOpen
  \bibfield  {author} {\bibinfo {author} {\bibfnamefont {T.~D.}\ \bibnamefont
  {Thomas}}, \bibinfo {author} {\bibfnamefont {L.~J.}\ \bibnamefont {Saethre}},
  \bibinfo {author} {\bibfnamefont {S.~L.}\ \bibnamefont {Sorensen}},\ and\
  \bibinfo {author} {\bibfnamefont {S.}~\bibnamefont {Svensson}},\ }\href
  {https://doi.org/10.1063/1.476646} {\bibfield  {journal} {\bibinfo  {journal}
  {J. Chem. Phys.}\ }\textbf {\bibinfo {volume} {109}},\ \bibinfo {pages}
  {1041} (\bibinfo {year} {1998})}\BibitemShut {NoStop}%
\bibitem [{\citenamefont {Jolly}\ and\ \citenamefont
  {Hendrickson}(1970)}]{jolly_thermodynamic_1970}%
  \BibitemOpen
  \bibfield  {author} {\bibinfo {author} {\bibfnamefont {W.~L.}\ \bibnamefont
  {Jolly}}\ and\ \bibinfo {author} {\bibfnamefont {D.~N.}\ \bibnamefont
  {Hendrickson}},\ }\href@noop {} {\bibfield  {journal} {\bibinfo  {journal}
  {J. Am. Chem. Soc.}\ }\textbf {\bibinfo {volume} {92}},\ \bibinfo {pages}
  {1863} (\bibinfo {year} {1970})}\BibitemShut {NoStop}%
\bibitem [{\citenamefont {Davis}\ and\ \citenamefont
  {Shirley}(1972)}]{davis_relaxation_1972}%
  \BibitemOpen
  \bibfield  {author} {\bibinfo {author} {\bibfnamefont {D.~W.}\ \bibnamefont
  {Davis}}\ and\ \bibinfo {author} {\bibfnamefont {D.~A.}\ \bibnamefont
  {Shirley}},\ }\href@noop {} {\bibfield  {journal} {\bibinfo  {journal} {Chem.
  Phys. Lett.}\ }\textbf {\bibinfo {volume} {15}},\ \bibinfo {pages} {185}
  (\bibinfo {year} {1972})}\BibitemShut {NoStop}%
\bibitem [{\citenamefont {Plashkevych}\ \emph {et~al.}(2000)\citenamefont
  {Plashkevych}, \citenamefont {Privalov}, \citenamefont {{\AA}gren},
  \citenamefont {Carravetta},\ and\ \citenamefont
  {Ruud}}]{plashkevych_validity_2000}%
  \BibitemOpen
  \bibfield  {author} {\bibinfo {author} {\bibfnamefont {O.}~\bibnamefont
  {Plashkevych}}, \bibinfo {author} {\bibfnamefont {T.}~\bibnamefont
  {Privalov}}, \bibinfo {author} {\bibfnamefont {H.}~\bibnamefont {{\AA}gren}},
  \bibinfo {author} {\bibfnamefont {V.}~\bibnamefont {Carravetta}},\ and\
  \bibinfo {author} {\bibfnamefont {K.}~\bibnamefont {Ruud}},\ }\href
  {https://doi.org/10.1016/S0301-0104(00)00171-3} {\bibfield  {journal}
  {\bibinfo  {journal} {Chem. Phys.}\ }\textbf {\bibinfo {volume} {260}},\
  \bibinfo {pages} {11} (\bibinfo {year} {2000})}\BibitemShut {NoStop}%
\bibitem [{\citenamefont {Becke}(1993)}]{becke_densityfunctional_1993}%
  \BibitemOpen
  \bibfield  {author} {\bibinfo {author} {\bibfnamefont {A.~D.}\ \bibnamefont
  {Becke}},\ }\href {https://doi.org/10.1063/1.464913} {\bibfield  {journal}
  {\bibinfo  {journal} {J. Chem. Phys.}\ }\textbf {\bibinfo {volume} {98}},\
  \bibinfo {pages} {5648} (\bibinfo {year} {1993})}\BibitemShut {NoStop}%
\bibitem [{\citenamefont {Stephens}\ \emph {et~al.}(1994)\citenamefont
  {Stephens}, \citenamefont {Devlin}, \citenamefont {Chabalowski},\ and\
  \citenamefont {Frisch}}]{stephens_ab_1994}%
  \BibitemOpen
  \bibfield  {author} {\bibinfo {author} {\bibfnamefont {P.~J.}\ \bibnamefont
  {Stephens}}, \bibinfo {author} {\bibfnamefont {F.~J.}\ \bibnamefont
  {Devlin}}, \bibinfo {author} {\bibfnamefont {C.~F.}\ \bibnamefont
  {Chabalowski}},\ and\ \bibinfo {author} {\bibfnamefont {M.~J.}\ \bibnamefont
  {Frisch}},\ }\href {https://doi.org/10.1021/j100096a001} {\bibfield
  {journal} {\bibinfo  {journal} {J. Phys. Chem.}\ }\textbf {\bibinfo {volume}
  {98}},\ \bibinfo {pages} {11623} (\bibinfo {year} {1994})}\BibitemShut
  {NoStop}%
\bibitem [{\citenamefont {Barca}\ \emph {et~al.}(2020)\citenamefont {Barca},
  \citenamefont {Bertoni}, \citenamefont {Carrington}, \citenamefont {Datta},
  \citenamefont {De~Silva}, \citenamefont {Deustua}, \citenamefont {Fedorov},
  \citenamefont {Gour}, \citenamefont {Gunina}, \citenamefont {Guidez},
  \citenamefont {Harville}, \citenamefont {Irle}, \citenamefont {Ivanic},
  \citenamefont {Kowalski}, \citenamefont {Leang}, \citenamefont {Li},
  \citenamefont {Li}, \citenamefont {Lutz}, \citenamefont {Magoulas},
  \citenamefont {Mato}, \citenamefont {Mironov}, \citenamefont {Nakata},
  \citenamefont {Pham}, \citenamefont {Piecuch}, \citenamefont {Poole},
  \citenamefont {Pruitt}, \citenamefont {Rendell}, \citenamefont {Roskop},
  \citenamefont {Ruedenberg}, \citenamefont {Sattasathuchana}, \citenamefont
  {Schmidt}, \citenamefont {Shen}, \citenamefont {Slipchenko}, \citenamefont
  {Sosonkina}, \citenamefont {Sundriyal}, \citenamefont {Tiwari}, \citenamefont
  {Galvez~Vallejo}, \citenamefont {Westheimer}, \citenamefont {Wloch},
  \citenamefont {Xu}, \citenamefont {Zahariev},\ and\ \citenamefont
  {Gordon}}]{GAMESS}%
  \BibitemOpen
  \bibfield  {author} {\bibinfo {author} {\bibfnamefont {G.~M.~J.}\
  \bibnamefont {Barca}}, \bibinfo {author} {\bibfnamefont {C.}~\bibnamefont
  {Bertoni}}, \bibinfo {author} {\bibfnamefont {L.}~\bibnamefont {Carrington}},
  \bibinfo {author} {\bibfnamefont {D.}~\bibnamefont {Datta}}, \bibinfo
  {author} {\bibfnamefont {N.}~\bibnamefont {De~Silva}}, \bibinfo {author}
  {\bibfnamefont {J.~E.}\ \bibnamefont {Deustua}}, \bibinfo {author}
  {\bibfnamefont {D.~G.}\ \bibnamefont {Fedorov}}, \bibinfo {author}
  {\bibfnamefont {J.~R.}\ \bibnamefont {Gour}}, \bibinfo {author}
  {\bibfnamefont {A.~O.}\ \bibnamefont {Gunina}}, \bibinfo {author}
  {\bibfnamefont {E.}~\bibnamefont {Guidez}}, \bibinfo {author} {\bibfnamefont
  {T.}~\bibnamefont {Harville}}, \bibinfo {author} {\bibfnamefont
  {S.}~\bibnamefont {Irle}}, \bibinfo {author} {\bibfnamefont {J.}~\bibnamefont
  {Ivanic}}, \bibinfo {author} {\bibfnamefont {K.}~\bibnamefont {Kowalski}},
  \bibinfo {author} {\bibfnamefont {S.~S.}\ \bibnamefont {Leang}}, \bibinfo
  {author} {\bibfnamefont {H.}~\bibnamefont {Li}}, \bibinfo {author}
  {\bibfnamefont {W.}~\bibnamefont {Li}}, \bibinfo {author} {\bibfnamefont
  {J.~J.}\ \bibnamefont {Lutz}}, \bibinfo {author} {\bibfnamefont
  {I.}~\bibnamefont {Magoulas}}, \bibinfo {author} {\bibfnamefont
  {J.}~\bibnamefont {Mato}}, \bibinfo {author} {\bibfnamefont {V.}~\bibnamefont
  {Mironov}}, \bibinfo {author} {\bibfnamefont {H.}~\bibnamefont {Nakata}},
  \bibinfo {author} {\bibfnamefont {B.~Q.}\ \bibnamefont {Pham}}, \bibinfo
  {author} {\bibfnamefont {P.}~\bibnamefont {Piecuch}}, \bibinfo {author}
  {\bibfnamefont {D.}~\bibnamefont {Poole}}, \bibinfo {author} {\bibfnamefont
  {S.~R.}\ \bibnamefont {Pruitt}}, \bibinfo {author} {\bibfnamefont {A.~P.}\
  \bibnamefont {Rendell}}, \bibinfo {author} {\bibfnamefont {L.~B.}\
  \bibnamefont {Roskop}}, \bibinfo {author} {\bibfnamefont {K.}~\bibnamefont
  {Ruedenberg}}, \bibinfo {author} {\bibfnamefont {T.}~\bibnamefont
  {Sattasathuchana}}, \bibinfo {author} {\bibfnamefont {M.~W.}\ \bibnamefont
  {Schmidt}}, \bibinfo {author} {\bibfnamefont {J.}~\bibnamefont {Shen}},
  \bibinfo {author} {\bibfnamefont {L.}~\bibnamefont {Slipchenko}}, \bibinfo
  {author} {\bibfnamefont {M.}~\bibnamefont {Sosonkina}}, \bibinfo {author}
  {\bibfnamefont {V.}~\bibnamefont {Sundriyal}}, \bibinfo {author}
  {\bibfnamefont {A.}~\bibnamefont {Tiwari}}, \bibinfo {author} {\bibfnamefont
  {J.~L.}\ \bibnamefont {Galvez~Vallejo}}, \bibinfo {author} {\bibfnamefont
  {B.}~\bibnamefont {Westheimer}}, \bibinfo {author} {\bibfnamefont
  {M.}~\bibnamefont {Wloch}}, \bibinfo {author} {\bibfnamefont
  {P.}~\bibnamefont {Xu}}, \bibinfo {author} {\bibfnamefont {F.}~\bibnamefont
  {Zahariev}},\ and\ \bibinfo {author} {\bibfnamefont {M.~S.}\ \bibnamefont
  {Gordon}},\ }\href {https://doi.org/10.1063/5.0005188} {\bibfield  {journal}
  {\bibinfo  {journal} {J. Chem. Phys.}\ }\textbf {\bibinfo {volume} {152}},\
  \bibinfo {pages} {154102} (\bibinfo {year} {2020})}\BibitemShut {NoStop}%
\bibitem [{\citenamefont {Kutzelnigg}\ \emph {et~al.}(1990)\citenamefont
  {Kutzelnigg}, \citenamefont {Fleischer},\ and\ \citenamefont
  {Schindler}}]{kutzelnigg1990iglo}%
  \BibitemOpen
  \bibfield  {author} {\bibinfo {author} {\bibfnamefont {W.}~\bibnamefont
  {Kutzelnigg}}, \bibinfo {author} {\bibfnamefont {U.}~\bibnamefont
  {Fleischer}},\ and\ \bibinfo {author} {\bibfnamefont {M.}~\bibnamefont
  {Schindler}},\ }in\ \href@noop {} {\emph {\bibinfo {booktitle} {Deuterium and
  shift calculation}}}\ (\bibinfo  {publisher} {Springer},\ \bibinfo {year}
  {1990})\ pp.\ \bibinfo {pages} {165--262}\BibitemShut {NoStop}%
\bibitem [{\citenamefont {Sakai}\ \emph {et~al.}(1997)\citenamefont {Sakai},
  \citenamefont {Miyoshi}, \citenamefont {Klobukowski},\ and\ \citenamefont
  {Huzinaga}}]{sakai_model_1997}%
  \BibitemOpen
  \bibfield  {author} {\bibinfo {author} {\bibfnamefont {Y.}~\bibnamefont
  {Sakai}}, \bibinfo {author} {\bibfnamefont {E.}~\bibnamefont {Miyoshi}},
  \bibinfo {author} {\bibfnamefont {M.}~\bibnamefont {Klobukowski}},\ and\
  \bibinfo {author} {\bibfnamefont {S.}~\bibnamefont {Huzinaga}},\ }\href@noop
  {} {\bibfield  {journal} {\bibinfo  {journal} {J. Chem. Phys.}\ }\textbf
  {\bibinfo {volume} {106}},\ \bibinfo {pages} {8084} (\bibinfo {year}
  {1997})}\BibitemShut {NoStop}%
\bibitem [{\citenamefont {Noro}\ \emph {et~al.}(1997)\citenamefont {Noro},
  \citenamefont {Sekiya},\ and\ \citenamefont {Koga}}]{noro_contracted_1997}%
  \BibitemOpen
  \bibfield  {author} {\bibinfo {author} {\bibfnamefont {T.}~\bibnamefont
  {Noro}}, \bibinfo {author} {\bibfnamefont {M.}~\bibnamefont {Sekiya}},\ and\
  \bibinfo {author} {\bibfnamefont {T.}~\bibnamefont {Koga}},\ }\href@noop {}
  {\bibfield  {journal} {\bibinfo  {journal} {Theor. Chem. Acc.}\ }\textbf
  {\bibinfo {volume} {98}},\ \bibinfo {pages} {25} (\bibinfo {year}
  {1997})}\BibitemShut {NoStop}%
\bibitem [{bsj()}]{bsjp}%
  \BibitemOpen
  \href@noop {} {}\bibinfo {note}
  {Http://sapporo.center.ims.ac.jp/sapporo/.}\BibitemShut {Stop}%
\bibitem [{\citenamefont {Tian}\ \emph {et~al.}()\citenamefont {Tian},
  \citenamefont {Duan}, \citenamefont {Hua},\ and\ \citenamefont
  {Luo}}]{DynaVib}%
  \BibitemOpen
  \bibfield  {author} {\bibinfo {author} {\bibfnamefont {G.}~\bibnamefont
  {Tian}}, \bibinfo {author} {\bibfnamefont {S.}~\bibnamefont {Duan}}, \bibinfo
  {author} {\bibfnamefont {W.}~\bibnamefont {Hua}},\ and\ \bibinfo {author}
  {\bibfnamefont {Y.}~\bibnamefont {Luo}},\ }\href@noop {} {\bibinfo {title}
  {\uppercase{D}yna\uppercase{V}ib, version 1.0}},\ \bibinfo {note}
  {\uppercase{R}oyal Institute of Technology: Sweden, 2012}\BibitemShut
  {NoStop}%
\bibitem [{\citenamefont {Ruhoff}(1994)}]{ruhoff_recursion_1994}%
  \BibitemOpen
  \bibfield  {author} {\bibinfo {author} {\bibfnamefont {P.~T.}\ \bibnamefont
  {Ruhoff}},\ }\href {https://doi.org/10.1016/0301-0104(94)00173-1} {\bibfield
  {journal} {\bibinfo  {journal} {Chem. Phys.}\ }\textbf {\bibinfo {volume}
  {186}},\ \bibinfo {pages} {355} (\bibinfo {year} {1994})}\BibitemShut
  {NoStop}%
\bibitem [{\citenamefont {Ruhoff}\ and\ \citenamefont
  {Ratner}(2000)}]{ruhoff_algorithms_2000}%
  \BibitemOpen
  \bibfield  {author} {\bibinfo {author} {\bibfnamefont {P.~T.}\ \bibnamefont
  {Ruhoff}}\ and\ \bibinfo {author} {\bibfnamefont {M.~A.}\ \bibnamefont
  {Ratner}},\ }\href
  {https://doi.org/10.1002/(SICI)1097-461X(2000)77:1<383::AID-QUA38>3.0.CO;2-0}
  {\bibfield  {journal} {\bibinfo  {journal} {Int. J. Quant. Chem.}\ }\textbf
  {\bibinfo {volume} {77}},\ \bibinfo {pages} {383} (\bibinfo {year}
  {2000})}\BibitemShut {NoStop}%
\bibitem [{\citenamefont {Nicolas}\ and\ \citenamefont
  {Miron}(2012)}]{nicolas_lifetime_2012}%
  \BibitemOpen
  \bibfield  {author} {\bibinfo {author} {\bibfnamefont {C.}~\bibnamefont
  {Nicolas}}\ and\ \bibinfo {author} {\bibfnamefont {C.}~\bibnamefont
  {Miron}},\ }\href {https://doi.org/10.1016/j.elspec.2012.05.008} {\bibfield
  {journal} {\bibinfo  {journal} {J. Electron. Spectrosc. Relat. Phenom.}\
  }\textbf {\bibinfo {volume} {185}},\ \bibinfo {pages} {267} (\bibinfo {year}
  {2012})}\BibitemShut {NoStop}%
\bibitem [{\citenamefont {Triguero}\ \emph {et~al.}(1999)\citenamefont
  {Triguero}, \citenamefont {Plashkevych}, \citenamefont {Pettersson},\ and\
  \citenamefont {Ågren}}]{triguero_separate_1999}%
  \BibitemOpen
  \bibfield  {author} {\bibinfo {author} {\bibfnamefont {L.}~\bibnamefont
  {Triguero}}, \bibinfo {author} {\bibfnamefont {O.}~\bibnamefont
  {Plashkevych}}, \bibinfo {author} {\bibfnamefont {L.}~\bibnamefont
  {Pettersson}},\ and\ \bibinfo {author} {\bibfnamefont {H.}~\bibnamefont
  {Ågren}},\ }\href {https://doi.org/10.1016/S0368-2048(99)00008-0} {\bibfield
   {journal} {\bibinfo  {journal} {J. Electron Spectros. Relat. Phenomena}\
  }\textbf {\bibinfo {volume} {104}},\ \bibinfo {pages} {195} (\bibinfo {year}
  {1999})}\BibitemShut {NoStop}%
\end{thebibliography}

%

\end{document}